\documentclass[12pt,preprint]{aastex}
\usepackage{graphicx}
\shorttitle{A triple stellar population in NGC\,6752} 
\shortauthors{A.\ P.\ Milone, et al.\ } 
\usepackage{ulem}

\begin{document}
\title{A WFC3/\textit{HST\/} view of the three stellar populations in the Globular Cluster NGC\,6752
          \footnote{           Based on observations with  the
                               NASA/ESA {\it Hubble Space Telescope},
                               obtained at  the Space Telescope Science
                               Institute,  which is operated by AURA, Inc.,
                               under NASA contract NAS 5-26555.}}
\author{
 A.\ P. \,Milone\altaffilmark{2,3,4}, 
 A.\ F. \,Marino\altaffilmark{2,5},
 G.\ Piotto\altaffilmark{6,7},
 L.\ R. \,Bedin\altaffilmark{7},
 J.\ Anderson\altaffilmark{8},  
 A.\ Aparicio\altaffilmark{3,4}, 
 A.\ Bellini\altaffilmark{8}, 
 S.\ Cassisi\altaffilmark{9}, 
 F.\ D'Antona\altaffilmark{10},  
 F.\ Grundahl\altaffilmark{11}, 
 M.\ Monelli\altaffilmark{3,4}, 
 D.\ Yong\altaffilmark{2} 
 } 

\altaffiltext{2}{Research School of Astronomy and Astrophysics, The Australian National University, Cotter Road, Weston, ACT, 2611, Australia; milone@mso.anu.edu.au, amarino@mso.anu.edu.au, david.yong@anu.edu.au}

\altaffiltext{3}{Instituto de Astrof\`\i sica de Canarias, E-38200 La Laguna, Tenerife, Canary Islands, Spain; aparicio@iac.es, monelli@iac.es}

\altaffiltext{4}{Department of Astrophysics, University of La Laguna, E-38200 La Laguna, Tenerife, Canary Islands, Spain}

\altaffiltext{5}{Max Planck Institute for Astrophysics, Postfach 1317, D-85741 Garching, Germany}

\altaffiltext{6}{Dipartimento  di   Astronomia,  Universit\`a  di Padova, Vicolo dell'Osservatorio 3, Padova I-35122, Italy; giampaolo.piotto@unipd.it }

\altaffiltext{7}{INAF-Osservatorio Astronomico di Padova, Vicolo dell'Osservatorio 5, I-35122 Padua, Italy; luigi.bedin@oapd.inaf.it}
 
\altaffiltext{8}{Space Telescope Science Institute, 3800 San Martin Drive, Baltimore, MD 21218; jayander@stsci.edu, bellini@stsci.edu}
 
\altaffiltext{9}{INAF-Osservatorio Astronomico di Collurania, via Mentore Maggini, I-64100 Teramo, Italy; cassisi@oa-teramo.inaf.it} 

\altaffiltext{10}{INAF-Osservatorio Astronomico di Roma, Via Frascati 33,
	   I-00040 Monte Porzio Catone, Rome, Italy; dantona@mporzio.astro.it}

\altaffiltext{11}{Department of Physics and Astronomy, Arhus University, Ny Munkegade, 8000, Arhus C, Denmark; fgj@phys.au.dk}

\begin{abstract}
Multi-band {\it Hubble Space Telescope} photometry reveals that the
 main sequence, sub-giant, and the red giant branch of the 
globular cluster NGC\,6752 splits into three main components 
in close analogy with the three distinct segments along its horizontal branch stars.
These triple sequences are consistent with three stellar groups: a stellar
population with a chemical composition similar to field halo stars (population a), a population (c) with enhanced sodium and nitrogen, depleted carbon
and oxygen and enhanced helium abundance ($\Delta Y \sim$0.03),
and a population (b) with an intermediate (between population a and c) chemical composition and slightly helium enhanced ($\Delta Y \sim$0.01).
These components contain $\sim$25\% (population a), $\sim$45\% (population b), and $\sim$30\% (population c) of the stars. 
No radial gradient for the relative numbers of the three populations has been identified out to about 2.5 half mass radii.
\end{abstract}


%%%%%%%%%%%%%%%%%%%%%%%%%%%%%%%%%%%%%%%%%%%%%%%%%%%%%%%
%
% SECTION 1 --- INTRODUCTION
%
%%%%%%%%%%%%%%%%%%%%%%%%%%%%%%%%%%%%%%%%%%%%%%%%%%%%%%%
%
\section{Introduction}
\label{introduction}
It is now widely accepted that many (if not all)
globular clusters (GCs) are made up of multiple populations of stars. 
Since the 70's, it was well known that stars within the same cluster have
light-element abundance variations
(e.g.\  Kraft 1979, Norris \& Freeman 1979), but, at that time it was not clear
whether these `abundance anomalies' were due to internal mixing or
differences in the primordial composition or a combination of these effects.  
More recently, high-resolution spectroscopy has revealed the presence of well-defined  correlations among some light-elements abundances (e.g.\ Kraft et al.\ 1992, Sneden et al.\ 1994, Ramirez \& Cohen 2002), including the  anticorrelations between Na and O, and Mg and Al, which indicate that material has been processed via high-temperature proton capture nucleosynthesis (Denisenkov \& Denisenkova 1989). 

The fact that the same light-elements variations have also been observed 
in unevolved cluster stars (e.g.\ Briley et al.\ 1994, 1997, Cannon et al.\ 1998, Gratton et al.\ 2001), 
whose internal temperatures do not allow high-T proton captures, 
and in fully-convective low-mass M-dwarfs (Milone et al.\ 2012) suggests that these stars were born with these chemical peculiarities imprinted in the matter from which they formed (Cottrell \& Da Costa 1981, see Gratton et al.\ 2004 for a review). 

High-precision {\it Hubble Space Telescope} (\textit{HST\/}) and
ground-based photometry has shown that several GCs host multiple main sequences (MSs)
including $\omega$ Centauri, NGC\,2808, 47 Tuc, NGC\,6752, and NGC\,6397
(Anderson 1997, Bedin et al.\ 2004, Piotto et al.\ 2007, Milone et al.\ 2010, 2012a,b), which have been associated
with stellar populations with different helium abundances
(D'Antona \& Caloi 2004, Norris 2004, Bedin et al.\ 2004, Piotto et al.\ 2005, D'Antona et al.\ 2005).  
Multiple stellar populations have also been detected in the color-magnitude diagram (CMD) from the presence of multiple sub-giant branches (SGBs, Milone et al.\ 2008, Anderson et al.\ 2009, Marino et al.\ 2012, Piotto et al.\ 2012) or multiple or spread red-giant branches (RGBs, e.g.\ Grundahl et al.\ 1998, 2000, Yong et al.\ 2008, Marino et al.\ 2008, Lee et al.\ 2009). 

Stellar evolution models predict that high-temperature H-burning
through the CNO cycle should result in an increase of the N abundance, at the expenses of C and O, and of an increase in the helium fraction.
Multiple stellar populations with different helium content could also account for the HB morphology of some GCs in which the bluer HB sequences can be associated with the presence of He-rich stars
(e.g.\ D'Antona et al.\ 2002, D'Antona \& Caloi 2008, Busso et al.\ 2007, Cassisi et al.\ 2009, Catelan, Valcarce \& Sweigart 2010, D'Alessandro et al.\ 2011). 
Indeed, clear evidence of the connection between the HB morphology 
with the multiple populations comes from Marino et al.\ (2011) who
have found that blue-HB stars of the GC M\,4 are all Na-rich and O-poor (hence He-rich), whereas red-HB stars are primarily Na-poor and O-rich (He-poor) (see also Norris 1981, Smith \& Norris 1993). Similar results have been found in NGC\,2808 (Gratton et al.\ 2011). 

This work adds yet another cluster (NGC\,6752) to the growing list of clusters with  photometric and spectroscopic evidence of multiple sequences along the RGB, MS, SGB, and HB.
We use  \textit{HST\/} filters covering a wide range of wavelengths to study the multiple stellar populations in the GC NGC\,6752. 
The presence of star-to-star light-element variations 
in the cluster has been widely reported in the literature (Norris et al.\ 1981, Grundahl et al.\ 2002, Yong et al.\ 2003, 2005, 2008, Carretta et al.\ 2007). 

Photometric evidence for three populations of stars along the RGB of NGC\,6752 with different Mg, Al, Mg, Si, Na, and O content was early identified by Grundahl et al.\ (2002) and Yong et al.\ 2008 (see also Carretta et al.\ 2012). These authors
found that the Str\"omgren photometric index $c_{1}$ correlates with
nitrogen abundance in stars both brighter and fainter than the RGB
bump, and suggest that the observed photometric scatter is due to stellar
populations with different N abundance (see also Milone et al.\ 2010, Kravtsov et al.\ 2011, Sbordone et al.\ 2011).

The HB of NGC\,6752 revealed also a complex structure, with two discontinuities that define three HB segments (Momany et al.\ 2002, 2004).
Villanova et al.\ (2009) analyzed spectra of seven HB stars with effective temperature $\sim$8000$<T_{\rm eff}<$9000 K and found that six of them have a chemical composition similar to field-halo stars, including helium.
A recent photometric analysis of data collected with the Advanced Camera for
Survey (ACS) on board  \textit{HST\/} 
also showed that NGC\,6752 has a broadened MS with some indication of a MS split (Milone et al.\ 2010) thus suggesting that its stellar populations also have different helium content. 

The paper is organized as follows. In Sect.~\ref{sec:data} we
present the data and describe the data reduction. In Sect.~\ref{sec:MS},
~\ref{sec:SGB},  and~\ref{sec:RGB} we study the triple MS, SGB,
 and RGB, respectively.
An effort was made to disentangle the 
multiple-populations  in each of these three evolutionary sequence separately.
In Sect.~\ref{sec:3msINT} we explore possible theoretical
 interpretations and estimate the helium difference among the three stellar populations. 
The study of the radial distributions of the various stellar populations is also undertaken. Finally, a summary and some discussion  
follow in Sect.~\ref{sec:discussion}.

\section{Observations and data reduction} 
\label{sec:data}
\begin{figure}[ht!]
\centering
\epsscale{.47}
\plotone{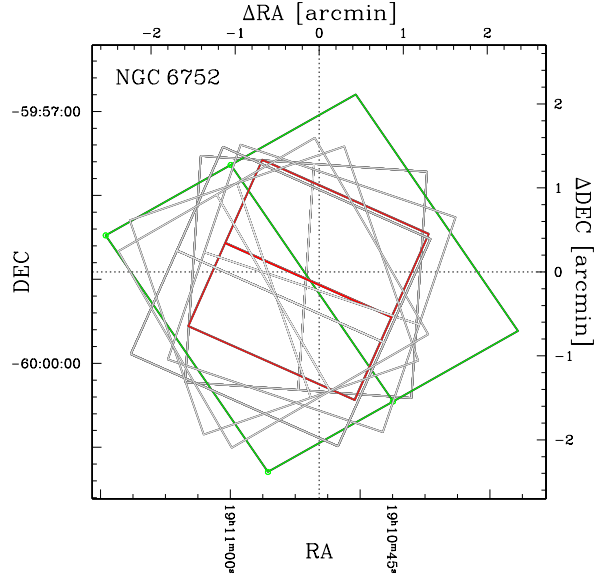}
%/home/milone/WORKS/SUMMARYNGC6752/FOOTPRINT/totnero.macro tot 2.7 NGC6752
\caption{Footprint of the {\it HST} fields studied in this paper. The footprint of ACS/WFC, WFC3/UVIS, and WFC3/NIR images are colored green, gray, and red, respectively. }
\label{footprint}
\end{figure}

This work makes use  of two data sets. 
For the central regions of the cluster we used 
both archival and proprietary material collected with
the UV-Visual (UVIS) and infra-red (IR) channels of the Wide Field
Camera 3 (WFC3), and the wide-field channel (WFC) of the Advanced Camera for Surveys (ACS) mounted at the  \textit{HST\/}.
Proprietary images (under program GO-12311, PI:\ Piotto) were
collected in two 1-orbit visits taken at two different orientations
the first one on March 23, and the second one on April 3, 2011, and consist of
12$\times$360 s images in camera/channel/filter WFC3/UVIS/F275W, and
2$\times$50 s in WFC3/UVIS/F814W.
We also used the photometric catalogs presented by Anderson et al.\ (2008) and obtained from ACS/WFC images taken under GO-10775 (PI:\ Sarajedini, see Sarajedini et al.\ 2007). 
The archive \textit{HST\/} material is described in table~\ref{tab:data} 
 and consists of images taken through 
fourteen filters spanning a wide spectral range, from the ultraviolet (F225W) to the infrared (F160W). Footprints of \textit{HST\/} images are shown in Fig.~\ref{footprint}. 

Photometry and relative positions of stars in \textit{HST\/}/WFC3 images were extracted with the software tools described in Bellini et al.\ (2011), mostly based on the software 
described in Anderson et al.\ (2006).
The photometry was calibrated onto the Vega-mag system following the
procedures given in Bedin et al.\ (2005), and using encircled energy
and zero points given at STScI's web pages.  
Star positions were corrected for geometric distortion using the solutions given by 
Bellini \& Bedin (2009) and Bellini et al.\ (2011) for WFC3/UVIS, Anderson \& King (2006) for ACS/WFC, 
and Anderson et al.\ (in preparation) for WFC3/NIR.

To study the external regions of the cluster, we made use of the ground-based photometric catalog published by Grundahl et al.\ (2002). 
 They have been obtained with the 1.54 m Danish telescope at La Silla (Chile) through the Str\"omgren filters $u,v,b,y$, and cover a field of view of $\sim$6$\times$6 arcmin centered on the cluster.
 They were reduced following the method outlined in some detail by Stetson (2005).

The stellar catalogs were purged of poorly measured objects
using quality indices that our software produces following the
procedure that it is described in Milone et al.\ (2009). 
Finally, the photometry was corrected for zero-point spatial variations 
following the recipes in Milone et al.\ (2012). 

%__________________________________________________________________

\begin{table}[!htp]
\center
\scriptsize {
\begin{tabular}{cccccl}
\hline
\hline
 INSTR &  DATE & N$\times$EXPTIME & FILTER  & PROGRAM & PI \\
%------------------------------------------------------------------------
\hline
WFC3/UVIS & Aug 21 2010 & 6$\times$120s                                             & F225W & 11904 & Kalirai \\
WFC3/UVIS & Mar 23, Apr 3 2011 & 12$\times$360                                      & F275W & 12311 & Piotto \\ 
WFC3/UVIS & May 05 2010 & 30s $+$ 2$\times$500s                                     & F336W & 11729 & Holtzman \\
WFC3/UVIS & May 05 2010 &  2$\times$2s $+$ 2$\times$348s$+$ $+$ 2$\times$880s       & F390W & 11664 & Brown \\
WFC3/UVIS & May 05 2010 & 50s $+$ 2$\times$700s                                     & F390M & 11729 & Holtzman \\
WFC3/UVIS & May 05 2010 & 90s $+$ 2$\times$1050s                                    & F395N & 11729 & Holtzman \\
WFC3/UVIS & May 05 2010 & 40s $+$ 2$\times$800s                                     & F410M & 11729 & Holtzman \\
WFC3/UVIS & May 05 2010 & 40s $+$ 2$\times$400s                                     & F467M & 11729 & Holtzman \\
WFC3/UVIS & Aug 7, 21 2010 & 12$\times$670s                                         & F502N & 11904 & Kalirai \\
WFC3/UVIS & May 05 2010 & 5s $+$ 40s $+$ 400s                                       & F547M & 11729 & Holtzman \\
WFC3/UVIS & Jul 31, Aug 7, 21 2010 & 15$\times$550s                                 & F555W & 11904 & Kalirai \\
WFC3/UVIS & May 01 2010 &  30s $+$ 2$\times$665s                                    & F555W & 11664 & Brown \\
WFC3/UVIS & Jul 31, Aug 7, 21 2010 & 15$\times$550s                                 & F814W & 11904 & Kalirai \\
WFC3/UVIS & May 01 2010 &  30s $+$ 2$\times$495s                                    & F814W & 11664 & Brown  \\
WFC3/UVIS & Mar 23, Apr 3 2011 & 2$\times$50                                        & F814W & 12311 & Piotto \\ 
WFC3/NIR   & May 01 2010 &  3$\times$4s $+$ 3$\times$49s $+$ 299s$+$ 2$\times$399s   & F110W & 11664 & Brown \\
WFC3/NIR   & May 01 2010 &  3$\times$4s $+$ 3$\times$49s $+$ 299s$+$ 2$\times$399s   & F160W & 11664 & Brown \\
ACS/WFC   & May 24 2006 & 1$\times$2s   + 4$\times$35s                              & F606W  & 10775 &  Sarajedini \\
ACS/WFC   & May 24 2006 & 1$\times$2s   + 4$\times$40s                              & F814W  & 10775 &  Sarajedini \\
%------------------------------------------------------------------------
\hline
\hline
\end{tabular}
}
\label{tab:data}
\caption{Description of the 
\textit{HST\/}
data sets used in this paper. }
\end{table}

\section{A triple main sequence in NGC\,6752}
\label{sec:MS}
%__________________________________________________________________
\begin{figure}[ht!]
\centering
\epsscale{.47}
\plotone{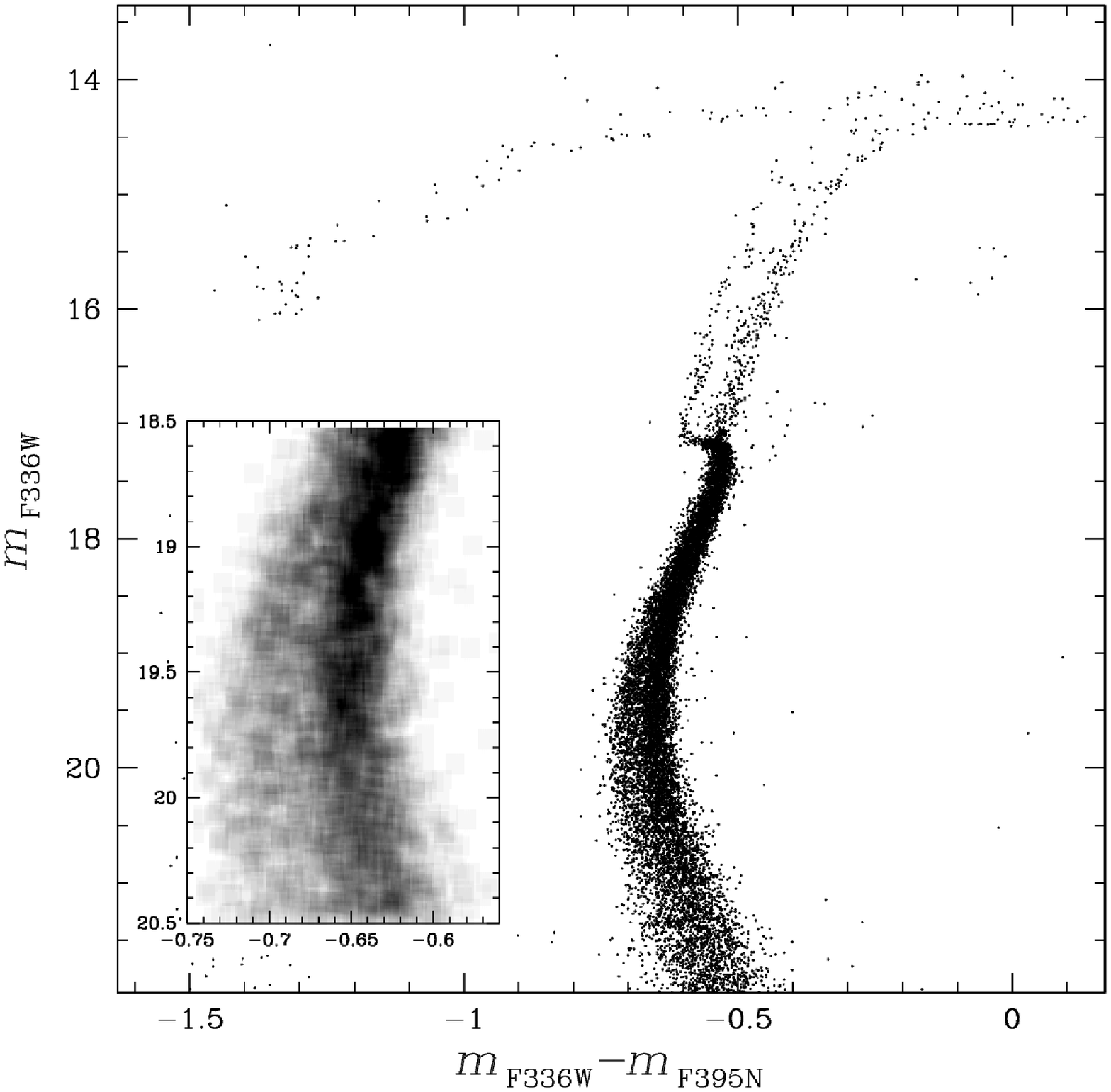}
\plotone{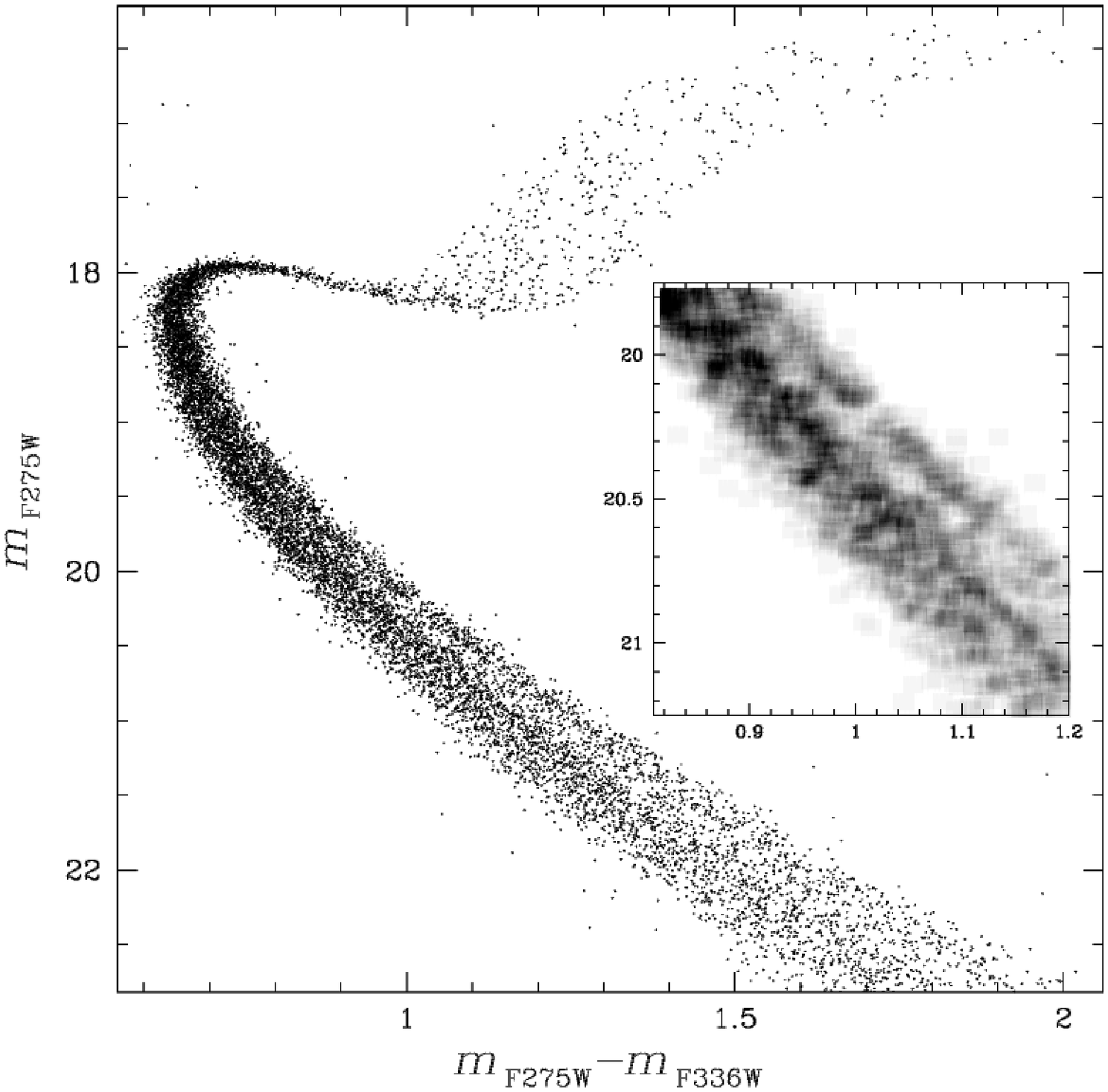}
%/home/milone/WORKS/SUMMARYNGC6752/MATCH336/FIGs/CMD.macro go go2
   \caption{$m_{\rm F336W}$ vs.\ $m_{\rm F336W}-m_{\rm F395N}$ (left
     panel) and $m_{\rm F275W}$ vs.\ $m_{\rm F275W}-m_{\rm F336W}$ CMDs
   of NGC\,6752.
The insets show the Hess diagram of a zoomed section of the MS.
}
\label{CMDs}
\end{figure}
%__________________________________________________________________
A visual inspection of the large number of CMDs that we obtain from
our data set confirms that  multiple sequences along
the MS and the RGB can be easily identified by using different combinations
of the F275W, F336W,
 and F395N filters. The left-hand panel of Fig.~\ref{CMDs} shows the 
$m_{\rm F336W}$ versus $m_{\rm F336W}-m_{\rm F395N}$ CMD after that
the photometric corrections and the quality selection described in the
previous Section were applied. We note a bimodal RGB, and a spread MS.
The $m_{\rm F275W}$ versus $m_{\rm F275W}-m_{\rm F336W}$ CMD plotted in
the right panel reveals an even larger number of features, with a
possible triple RGB, and a clear split MS composed of two distinct
components. A narrow red MS, containing about 
one fourth of MS stars,
and a more dispersed blue MS.  

In recent papers, we have shown that two-color diagrams obtained by combining a far-UV filter
(such as F225W or F275W), a near-UV filter (such as F336W) and a
visible filter (such as F438W) are powerful way
for identifying
populations of stars with different helium and light-element
abundances (see Milone et al.\ 2012a,b for results on 47\,Tuc and NGC\,6397). 

\begin{figure}[ht!]
\centering
\epsscale{.75}
\plotone{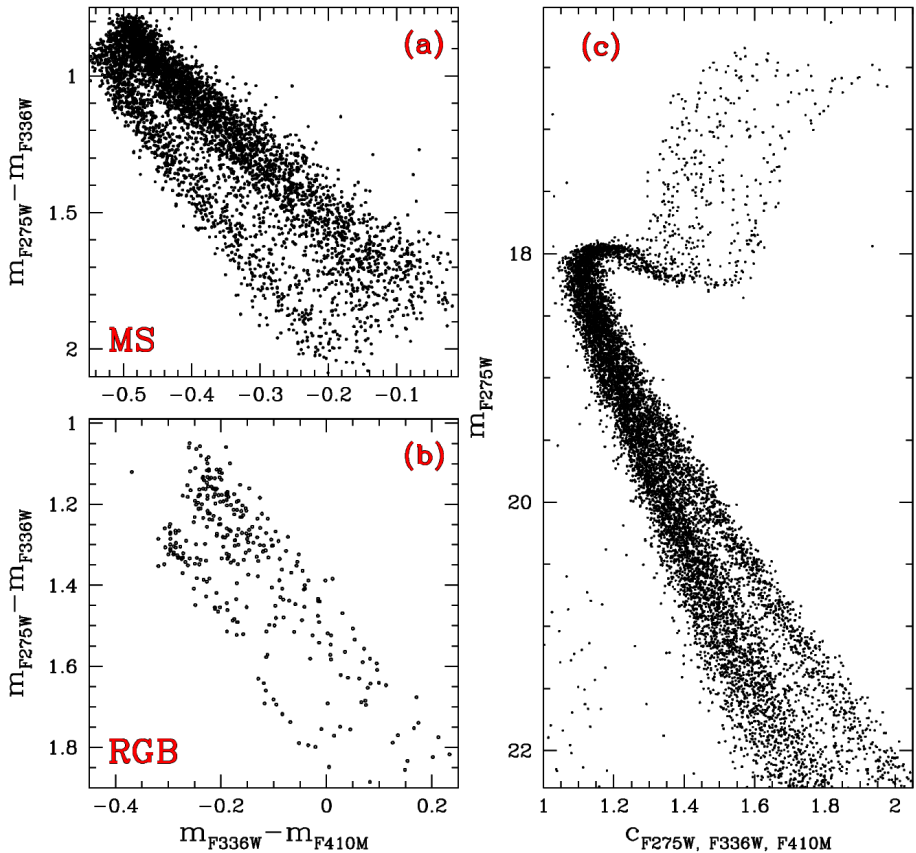}
%/home/milone/WORKS/SUMMARYNGC6752/MATCH336/FIGs/CMD.macro go3 old
%/home/milone/WORKS/SUMMARYNGC6752/MATCH336/ZP4Mf/fig.macro go1
   \caption{$m_{\rm F275W}-m_{\rm F336W}$ versus $m_{\rm F336W}-m_{\rm F410M}$ two color diagram  for MS stars with $19.65<m_{\rm F275W}<23.25$ (panel (a)), and for RGB stars with $m_{\rm F275W}<18.0$
(panel (b)). \textit{Panel (c)}: The
$m_{\rm F275W}$ versus $c_{\rm F275W, F336W, F410M}$ diagram for all NGC\,6752 stars
in our sample.   
}
\label{275336390}
\end{figure}

Motivated by these results, in Fig.~\ref{275336390}a we plot $m_{\rm F275W}-m_{\rm F336W}$ against $m_{\rm F336W}-m_{\rm F410M}$ for MS stars with
$19.65<m_{\rm F275W}<23.25$.  Panel b shows the same 
two-color diagram
for the RGB stars.

We also defined the color index $c_{\rm F275W, F336W, F410M}$=$(m_{\rm F275W}-m_{\rm F336W}) - (m_{\rm F336W}-m_{\rm F410M})$. Quite interestingly, the $m_{\rm F275W}$ versus $c_{\rm F275W, F336W, F410M}$ color-index-Magnitude Diagram (CMD) 
of Fig.~\ref{275336390}c allows us to identify multiple sequences along the 
entire CMD, from the MS to the RGB tip.
There is a clear color spread, with the presence of three RGBs and two distinct MSs, in close analogy with  what was observed in 47\,Tuc and NGC\,6397. 
Also the SGB is not consistent with a simple stellar population.
In the following we will refer to 
the less-populated MS located on the bottom-left of the two-color diagram, 
in Fig.~2a, 
as MSa.  
The MS is analyzed in this Section, while Sects.~\ref{sec:SGB} and~\ref{sec:RGB} are dedicated to the SGB and the RGB. 

%__________________________________________________________________
%__________________________________________________________________

In Milone et al.\ (2010),  we used high-precision photometry of 
 ACS/\textit{HST\/} images to search for signs of multiple
populations in NGC\,6752. We found a broadened MS, and demonstrated that
this broadening is intrinsic. We  also noted a possible 
MS bi-modality, and suggested that the MS split could be due to two
stellar populations with almost the same age and iron abundance, but 
 different helium content.
As demonstrated in the following, 
the data set presented in this paper allows us 
to identify multiple populations with different helium content
with an higher accuracy than was possible for Milone et al.\  (2010), with the
data available at that time.

\begin{figure}[ht!]
\centering
\epsscale{.6}
\plotone{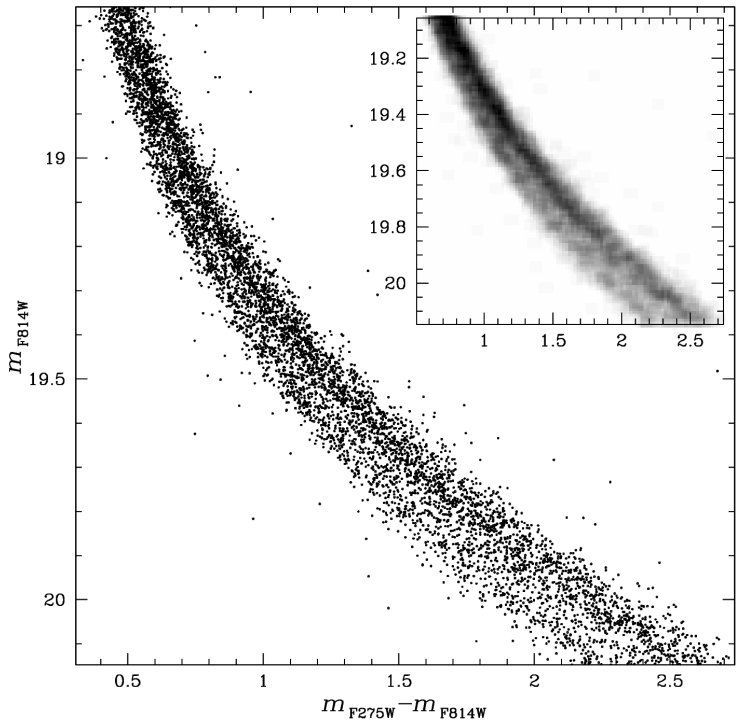}
%/home/milone/WORKS/SUMMARYNGC6752/MATCH814/FIGs/msratio.macro go3
   \caption{$m_{\rm F814W}$ vs.\ $m_{\rm F275W}-m_{\rm F814W}$ CMD for
   MS stars of NGC\,6752. The Hess diagram in the inset is a zoom of
   the region where the MS bimodality is more evident.}
\label{F275F814ms}
\end{figure}

Both theoretical arguments and observations indicate that CMDs 
  with wide color baselines can be very sensitive to helium differences among
 stars (e.g.\ D'Antona et al.\ 2005, Piotto et al.\ 2007).
In this context, the $m_{\rm F814W}$ vs.\ $m_{\rm  F275W}-m_{\rm F814W}$ CMD shown in Fig.~\ref{F275F814ms} clearly reveals a bimodal MS. 
The MS bi-modality is even  more evident in the Hess diagram plotted in
the inset. The two MSs  merge close to the turn off, and the MS
separation increases towards fainter magnitudes, from about 0.1 mag at
$m_{\rm F814W}$$\sim$19.05  up to 0.25 mag at $m_{\rm  F814W}$$\sim$20.15. Hereafter, we will refer to  the bluest MS of
Fig.~\ref{F275F814ms} as MSc. We will also demonstrate that the MSa
and the MSc correspond to different stellar populations.
Note the different morphology (i.e.\ different distribution in color of the
stars) of the MSs plotted in Fig.~\ref{F275F814ms} and Fig.~\ref{CMDs},
right panel.

%__________________________________________________________________
%__________________________________________________________________

\begin{figure}[ht!]
\centering
\epsscale{.75}
\plotone{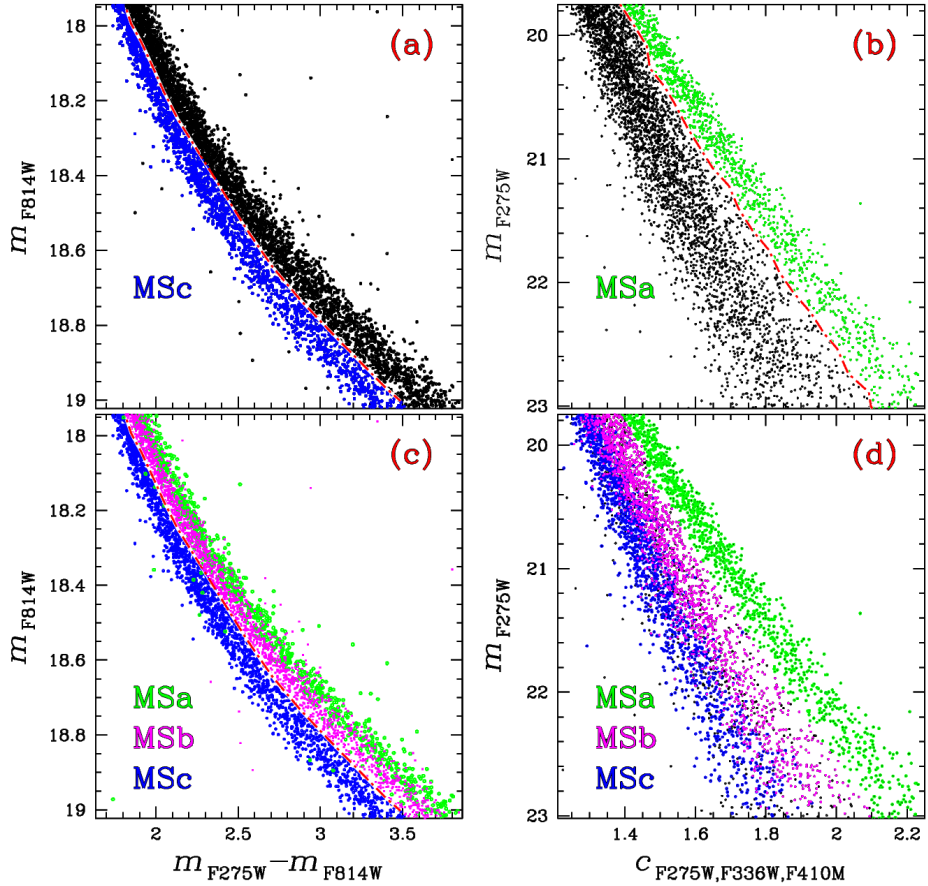}
%/home/milone/WORKS/SUMMARYNGC6752/MATCH814/FIGs/FIG.macro go go2
   \caption{Visualization of the procedure to define the
   sample of MSa, MSb, and MSc stars. The red dashed-dotted lines in
   panels (a) and (b) are used to select MSc and MSa stars 
   that we colored blue and green respectively. 
  In the panels (c) and (d) we adopted a magenta color to identify
   stars that belong neither to the MSa nor to the MSc, and
   hence are part of the third MS component (MSb). }
\label{selMS}
\end{figure}

In order to compare the sequences identified in this `regular $m_{\rm  F814W}$ vs.\ $m_{\rm F275W}-m_{\rm F814W}$ CMD' with those identified in the $m_{\rm F275W}$ 
\textit{vs.\ }  $c_{\rm F275W, F336W, F410M}$ CMD of Fig.~2c, 
we color-code the MSa stars green, and in blue the MSc stars 
(see panels (a) and (b) of Fig.~\ref{selMS}).  
Panel (c) of Fig.~\ref{selMS} shows that MSa stars are distinct from MSc, and that MSa$+$MSc stars are not all MS stars in NGC\,6752.
In the bottom panels of Fig.~\ref{selMS}, we identify the MSa and MSc stars
in {\it both} panels, and identify as `MSb' stars 
those that are neither MSa or MSc.  These stars are colored magenta.

This approach allowed us to demonstrate that the MSa and the MSc correspond
to two distinct stellar populations, and that NGC\,6752 hosts a 
at least a third MS population, labeled as MSb. 
%
%__________________________________________________________________
%__________________________________________________________________

\begin{figure}[ht!]
\centering
\epsscale{.75}
\plotone{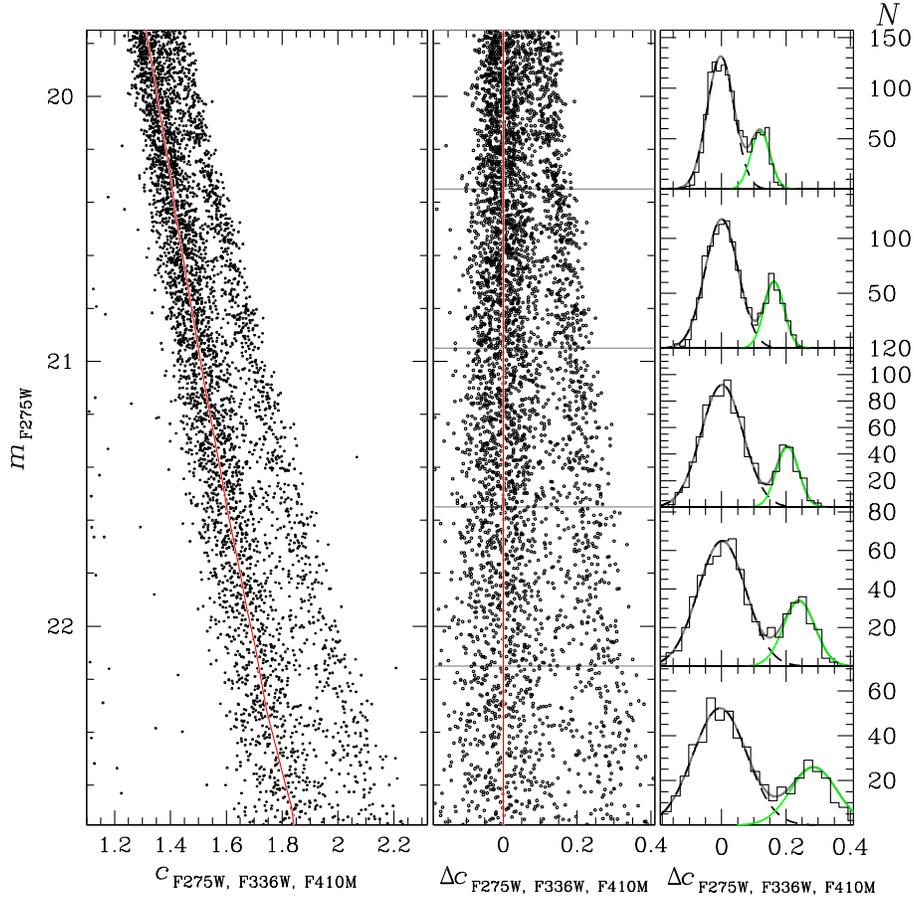}
%/home/milone/WORKS/SUMMARYNGC6752/MATCH336/FIGs/msratio.macro
%/home/milone/WORKS/SUMMARYNGC6752/MATCH336/ZP4Mf/msratio.macro
   \caption{\textit{Left panel:} Reproduction of the diagram of
     Fig.~\ref{275336390}c, the red line is the fiducial of the most
     populous MS. \textit{Middle panel:}
     The same diagram, after subtraction of the color of
     the fiducial line. \textit{Right panel:} The 
$\Delta c_{\rm F275W, F336W, F410M}$ color distribution in five $m_{\rm F275W}$ intervals. The gray lines represent the least-square best fits  of two Gaussians to the observed distribution.} 
\label{MSaRatio}
\end{figure}

%%%
The method used to estimate the fraction of stars in the MSa is
illustrated in Fig.~\ref{MSaRatio}. The left-hand panel of the figure
is a reproduction of the CMD 
of Fig.~\ref{275336390}c
with the fiducial line of the most populated MS superimposed. 
The verticalized MS is plotted in the middle panel and the right-hand panels 
show the  histograms of the $\Delta (c_{\rm F275W, F336W, F410M})$ color distribution in five $m_{\rm F275W}$ intervals. We fitted each histogram with
the sum of two Gaussians, colored green and black. Hereafter, the green color code will be used to highlight MSa stars. From the area under
the Gaussians we estimate that 25$\pm$2\% of stars belongs to MSa. 
The errors were computed as the rms of the values obtained for the five bins, and then divided by two
(i.e.\, the square-root of the number of bins minus one).

%__________________________________________________________________
%__________________________________________________________________

The procedure used to obtain the fiducial line and to verticalize the MS 
has been adopted in several previous papers from our group 
(e.g.\ Piotto et al.\ 2007), and will be used 
several times in the remaining part 
of this paper. 
Briefly, we adopt the most populous (in this CMD, the bluest) MS as reference
sequence and draw by hand a first approximation ridge line.
We also select a color range around this line to include most of the
stars on the blue MS. 
Then we divide the reference sequence in 0.15 mag
intervals, and calculate the median colors of the stars within each interval.
These median points are then interpolated with a spline.
We calculate the  spline at the magnitude level of each star and subtracted it
 from each star's color to estimate the MSRL residual for each.  We then
 determined a sigma-clipped mean for each magnitude interval, and repeated
   the procedure several times.  
The result is a fiducial line plotted in Fig.~\ref{MSaRatio}. Finally
we subtract from the $c_{\rm F275W, F336W, F410M}$ color of each star
the corresponding color on the fiducial line at the same $m_{\rm F275W}$.

\begin{figure}[ht!]
\centering
\epsscale{.75}
\plotone{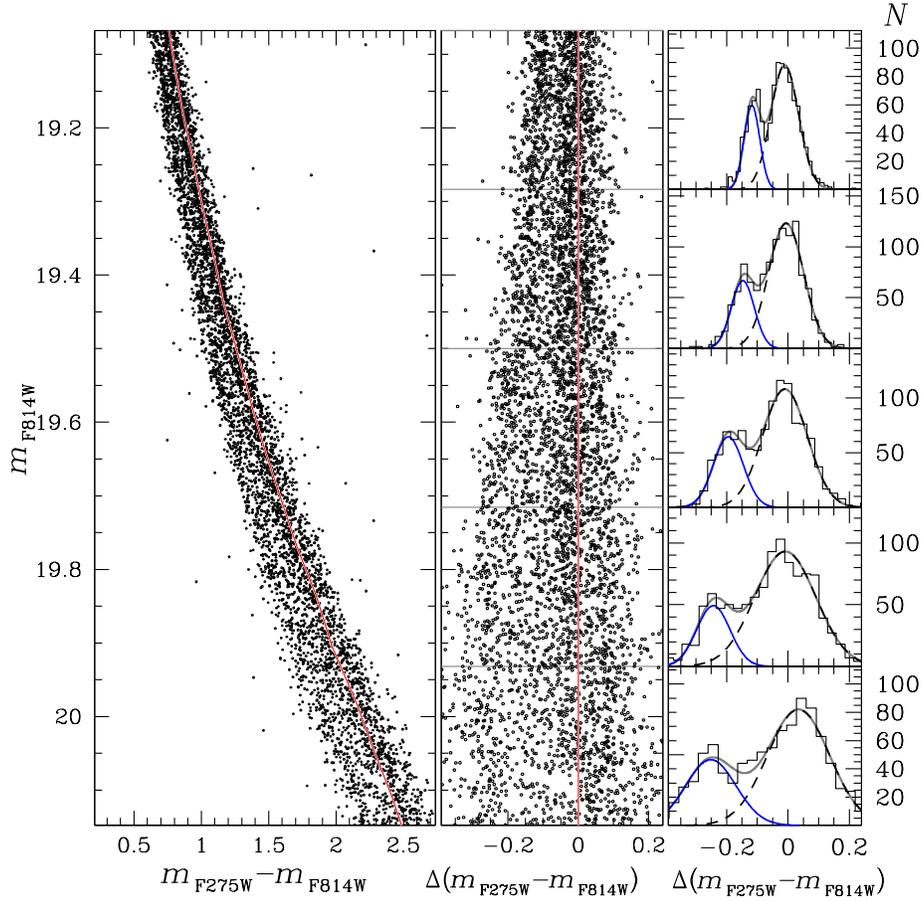}
%/home/milone/WORKS/SUMMARYNGC6752/MATCH814/FIGs/msratio.macro
\caption{$m_{\rm F814W}$ vs.\ $m_{\rm  F275W}-m_{\rm F814W}$ CMD
and verticalized $m_{\rm F814W}$ vs.\ $\Delta$($m_{\rm F275W}-m_{\rm F814W}$) diagram 
(left and middle panels). Right panels show the histogram of the color distribution 
in five $m_{\rm F814W}$ intervals. Gray lines represent the least-square best fits 
of two Gaussians to the observed distribution. }
\label{MSRATIOc}
\end{figure}

%%%%%%%%%%%
In order to calculate the fraction of MSc stars we followed a recipe similar
to the one already applied to the MSa, and illustrated in
Fig.~\ref{MSRATIOc}. In the left-hand panel we plotted the  $m_{\rm F814W}$
 vs.\ $m_{\rm  F275W}-m_{\rm F814W}$ CMD, while, in the central panel,
 we show the verticalized CMD, obtained with the same procedure as
 explained above. The color distribution histograms for five magnitude intervals
 are shown in the right-hand panels and are fitted with two
 Gaussians colored blue and black. In the following, the blue color will be used
 to indicate MSc stars. From the area under the best fit Gaussians, we
 estimate that 31$\pm$3\% of the total number of MS stars belong to
MSc. Errors are calculated as described above for the case of the MSa.
Since we have already estimated that the MSa and the MSc contain 
25$\pm$2\% and 31$\pm$3\% of the total number of MS stars, we can
conclude that the MSb is made up of the remaining 44$\pm$4\% of MS stars. 
%__________________________________________________________________
%__________________________________________________________________

\begin{figure}[ht!]
\centering
\epsscale{.75}
\plotone{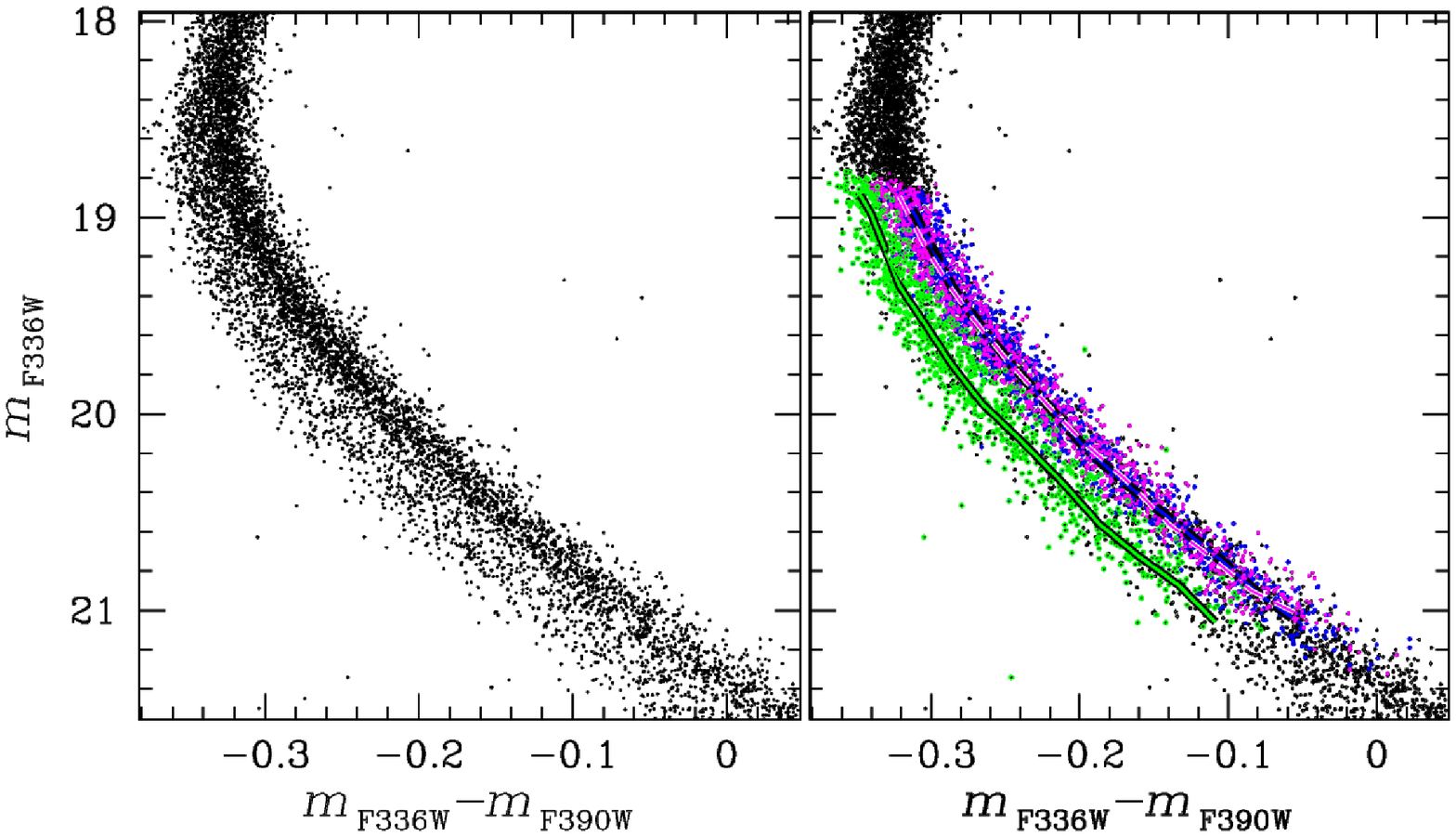}
%/home/milone/WORKS/SUMMARYNGC6752/MATCH814/FIGs/CMD.macro go4 go5
   \caption{\textit{Left panel:} $m_{\rm F336W}$ versus
$m_{\rm F336W}-m_{\rm F390W}$ CMD for MS stars. Right panel shows an
     example of the definition of the fiducial lines. The sample of
     MSa, MSb, and MSc stars defined in Fig.~\ref{selMS} are colored
     green with black shadow, magenta  with white shadow, and blue   with black shadow, respectively. In this case, MSb and MSc fiducials are largely overlapping. The same color codes 
     adopted in the previous figures are also
     used to represent the corresponding fiducials.} 
\label{fiducialMS}
\end{figure}

With these identifications and the large number of filters through
 which we have observations, we can analyze the relative location
of the three MSs in a large number of CMDs. As an example,  
left-panel of Fig.~\ref{fiducialMS} shows the $m_{\rm F336W}$ versus
$m_{\rm F336W}-m_{\rm F390W}$ CMD for MS stars. In the right-panel
we plotted the same CMD, but using the same color codes previously
defined in order to highlight MSa, MSb, and MSc stars. Contrary
 to what is observed in the $m_{\rm  F275W}-m_{\rm F814W}$ color in Fig.~3 (or in Fig.~4), in this CMD the MSa is bluer than the
bulk of MS stars, with the MSb and the MSc being almost coincident.
%__________________________________________________________________
%__________________________________________________________________

\begin{figure}[ht!]
\centering
\epsscale{.95}
\plotone{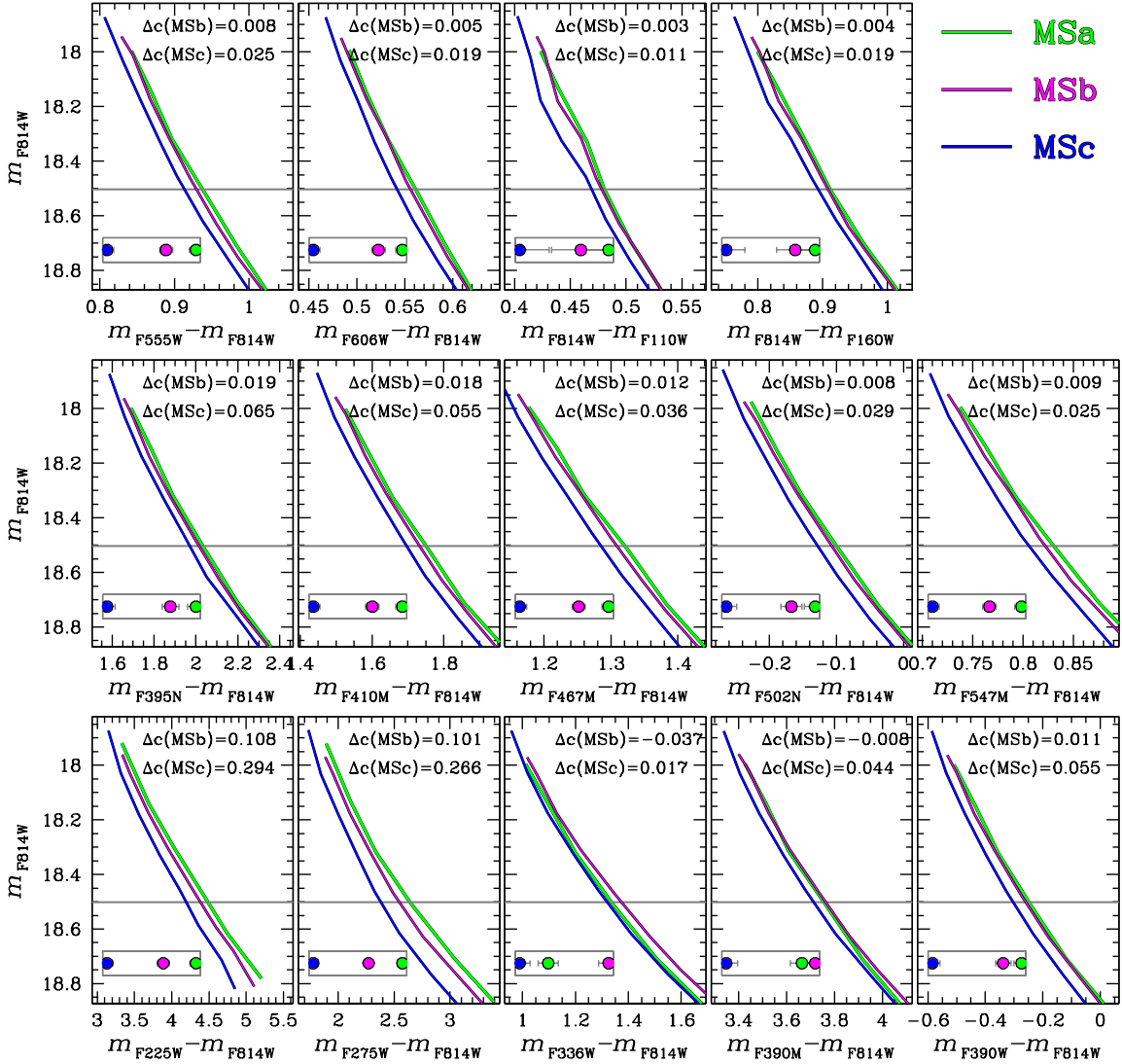}
%/home/milone/WORKS/SUMMARYNGC6752/MATCH814/FIGs/MSRLn.macro gor go
   \caption{MS fiducials in fourteen $m_{\rm F814W}$ versus $m_{\rm X}-m_{\rm F814W}$ (or $m_{\rm  F814W}-m_{\rm X}$) CMDs 
(X=F225W, F275W, F336W, F390M, F390W, F395N, F410M, F467M, F502N,
F547M, F555W, F606W, F110W, and F160W). At the top of each
panel we give the color distance from the MSa of the other two MSs, measured at 
$m_{\rm  F814W}^{\rm cut}$=18.5.
The positions of MSa, MSb, and MSc at $m_{\rm  F814W}^{\rm cut}$=18.5
are represented with green, magenta, and blue circles, respectively in the inset of each CMD.
}
\label{FIDUCIALs}
\end{figure}

In order to follow the behaviour of the three MSs in all the possible combinations of the
photometric bands available in our data set, 
we followed the approach illustrated in Fig.~\ref{FIDUCIALs}. 
In each panel we show the fiducial lines of the three MSs in the
$m_{\rm F814W}$ versus $m_{\rm X}-m_{\rm F814W}$ (or $m_{\rm
  F814W}-m_{\rm X}$) plane, where 
X=F225W, F275W, F336W, F390M, F390W, F395N, F410M, F467M, F502N,
F547M, F555W, F606W, F110W, and F160W. 
 For the two IR filters, F110W and F160W, we 
adopted the $m_{\rm F814W}-m_{\rm X}$ color baseline, 
being in these cases X the redder filter.

The MSs color properties in the various CMDs can be summarized as follows:\\

i) The MSb is typically bluer than the MSa in all CMDs of Fig ~\ref{FIDUCIALs},
with the exception of the  $m_{\rm F814W}$ vs.\ $m_{\rm F336W}-m_{\rm F814W}$ and the $m_{\rm
  F814W}$ vs.\ $m_{\rm F390M}-m_{\rm F814W}$ CMDs, where the two sequences
invert their relative colors. \\

ii) The MSc is bluer than the MSa in all CMDs.\\
 
iii) The color distance between the MSa and both the MSb and the MSc
increases for larger color baselines, with the exception of the $m_{\rm
  F336W}-m_{\rm F814W}$, the $m_{\rm F390M}-m_{\rm F814W}$,  and the
$m_{\rm F390W}-m_{\rm F814W}$ colors.\\
 
iv) The color separation between the MSc and the MSb increases with the
size of the color baseline for all colors studied in this paper.\\

In the following, we will use these data to gather information on the chemical
composition of the three MSs.
%
%__________________________________________________________________
%__________________________________________________________________

\section{Multiple stellar populations along the Sub Giant Branch}
\label{sec:SGB}
The first photometric evidence of multiple stellar populations along the SGB of NGC\,6752 
%AM: F
comes 
from the recent paper by Kravtsov et al.\ (2011).  Using wide-field
ground-based photometry, Kravtsov et al.\ (2011) identified a spread of $\sim$0.3
magnitudes in $U$ band, with the faintest SGB more centrally
concentrated than its brighter counterpart.

\begin{figure}[ht!]
\centering
\epsscale{.75}
\plotone{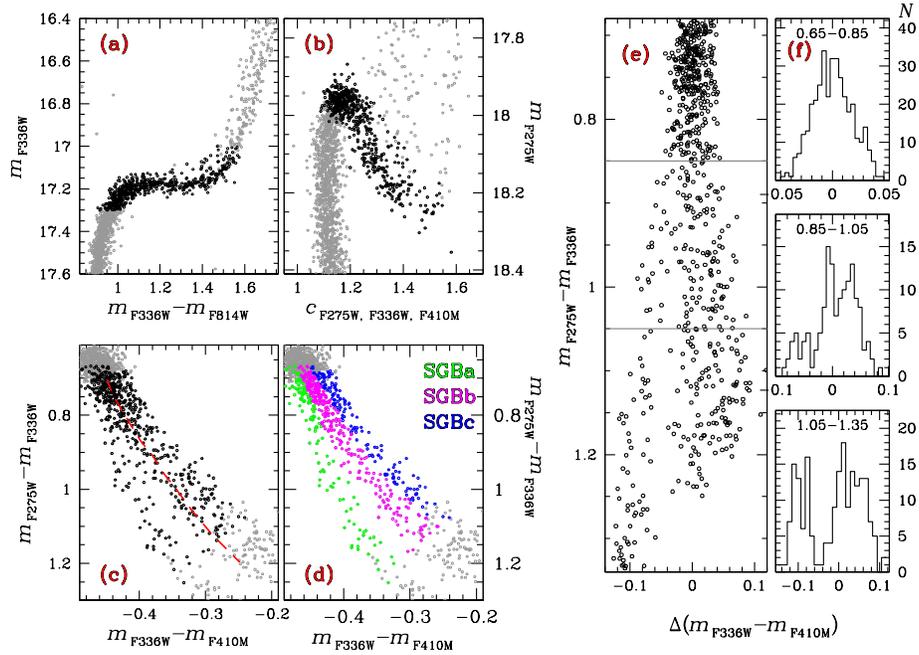}
%/home/milone/WORKS/SUMMARYNGC6752/MATCH336/FIGs/CMD.macro hi
   \caption{$m_{\rm F336W}$ versus $m_{\rm F336W}-m_{\rm
  F814W}$ CMD (panel a) and $m_{\rm F275W}$ versus $c_{\rm F275W, F336W,
  F410M}$ diagram (panel b) zoomed around the SGB. SGB stars are 
highlighted in thick 
black. \textit{Panel (c):} $m_{\rm F275W}-m_{\rm F336W}$ versus
$m_{\rm F336W}-m_{\rm F410M}$ two-color diagrams for the stars shown
in the upper panels. 
The red-dashed line is a fiducial line drawn by hand through the middle SGB.
In the panel (d), three groups of SGBa, SGBb, and SGBc stars are 
defined and color-coded in green, magenta, and blue, respectively.
The verticalized $m_{\rm F275W}-m_{\rm F336W}$ versus $\Delta(m_{\rm F336W}-m_{\rm F410M}$) is plotted in panel (e), while the histogram 
of the distribution in $\Delta(m_{\rm F336W}-m_{\rm F410M}$) 
is shown in panels (f) for the three quoted $m_{\rm F275W}-m_{\rm F336W}$ intervals.
}
\label{seleSGB}
\end{figure}

Our multi-color set of CMDs reveals an even more complex picture for the SGB of NGC\,6752. 
A visual inspection at the $m_{\rm F336W}$ versus $m_{\rm F336W}-m_{\rm F814W}$ CMD and the
$m_{\rm F275W}$ versus $c_{\rm F275W, F336W, F410M}$ diagram in the upper panel of Fig.~\ref{seleSGB} immediately reveals that, in the WFC3/UVIS field of view, there is {\it no} evidence for a wide
magnitude spread along the SGB in F336W (which is the \textit{HST\/} analog of the standard $U$), even though the SGB is clearly not
consistent with a single stellar population. By analogy with what we
did for the MS stars, in the lower panels of Fig.~\ref{seleSGB}, we
plot the $m_{\rm F275W}-m_{\rm F336W}$ versus $m_{\rm F336W}-m_{\rm F410M}$ 
two-color diagram. This diagram shows a multi-modal distribution of SGB stars
(see also the histrograms of panels f). 
As shown
in the figure, we selected by eye three groups of SGB stars that we named SGBa, SGBb, and SGBc and
colored green, magenta, and blue, respectively.
%__________________________________________________________________
%__________________________________________________________________

\begin{figure}[ht!]
\centering
\epsscale{.95}
\plotone{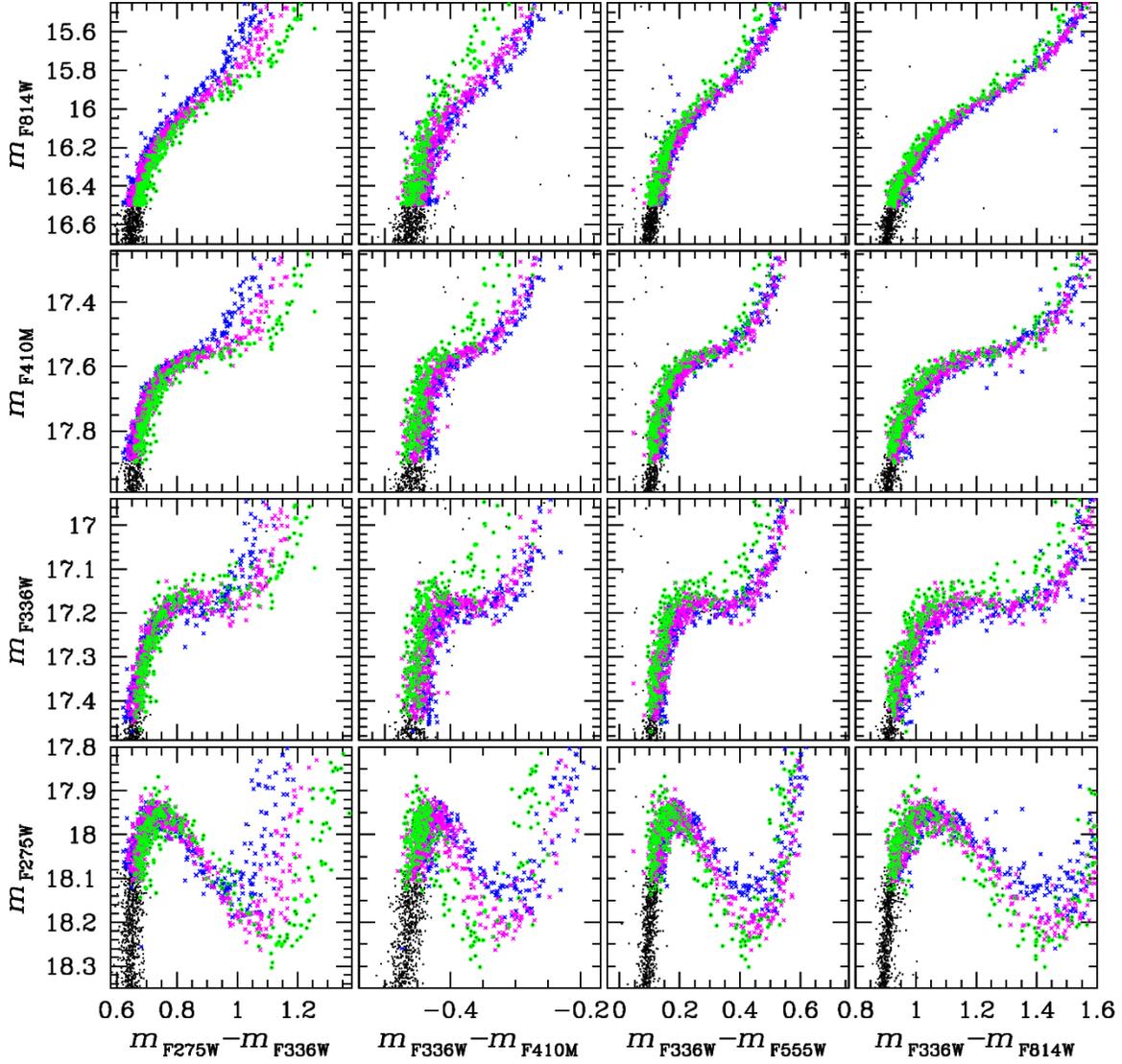}
%/home/milone/WORKS/SUMMARYNGC6752/MATCH336/FIGs/sgbs.macro
\caption{Collection of CMDs zoomed around the SGB. SGBa, SGBb, and
  SGBc stars defined in 
Fig.~\ref{seleSGB}d are plotted green, magenta,
  and blue, respectively.  }
\label{sgbs}
\end{figure}

In order to better understand the properties of these three sequences,
Fig.~\ref{sgbs} gives a 4 $\times$ 4 array of CMDs, where stars of the
three sequences  selected in Fig.~\ref{seleSGB} are plotted with their
color code.  
Figure~\ref{sgbs} shows some significant features of
the SGB of NGC\,6752:

(i) SGBa stars share some similarities with MSa:\ 
   they are on average redder than the bulk of SGB stars
   in $m_{\rm F275W}$-$m_{\rm F336W}$, but they become bluer than the remaining 
   SGB stars in the other CMDs of the first row of Fig.~\ref{sgbs}.
   
(ii) In the F336W band, SGBa stars are typically brighter 
than the other SGB stars. This fact also explains why they appear redder than SGBb 
and SGBc stars in $m_{\rm F275W}-m_{\rm F336W}$, and bluer than
   them in the other CMDs of the third row of Fig.~\ref{sgbs}.
 
(iii) In all the CMDs, the SGBb sequence seems to be located between the SGBa and the SGBc.

The analogy in color distribution of the three MSs and SGBs justifies
the names that we gave to these sequences, which explicitly want to suggest that the SGBa, 
SGBb, and SGBc represent the continuation along the CMD of the
MSa,  MSb, and MSc, respectively.
%__________________________________________________________________
%__________________________________________________________________

\section{Multiple stellar populations along the Red Giant Branch}
\label{sec:RGB}
As already mentioned in Sect.~\ref{introduction}, the first evidence
that the RGB of NGC\,6752 is not consistent with a single stellar
population comes from Grundahl et al.\ (2002) 
  followed by Yong et al.\ (2008),  Milone et al.\ (2010), Kravtsov et al.\ (2011), and Carretta et al.\ (2012). These studies have detected a large spread in
in the $c_{\rm y}$ Str\"omgren index\footnote{
The index $c_{\rm y}$ is defined as  $c_{\rm y}$=$c_{1}-(b-y)$, where $c_{1}$=$(b-v)-(v-b)$ (Yong et al.\ 2008).
} 
with the presence of three RGBs. The large data set listed in
Table~1 allowed us to add further information on the
multimodal RGB of NGC\,6752.

\begin{figure}[ht!]
\centering
\epsscale{.85}
\plotone{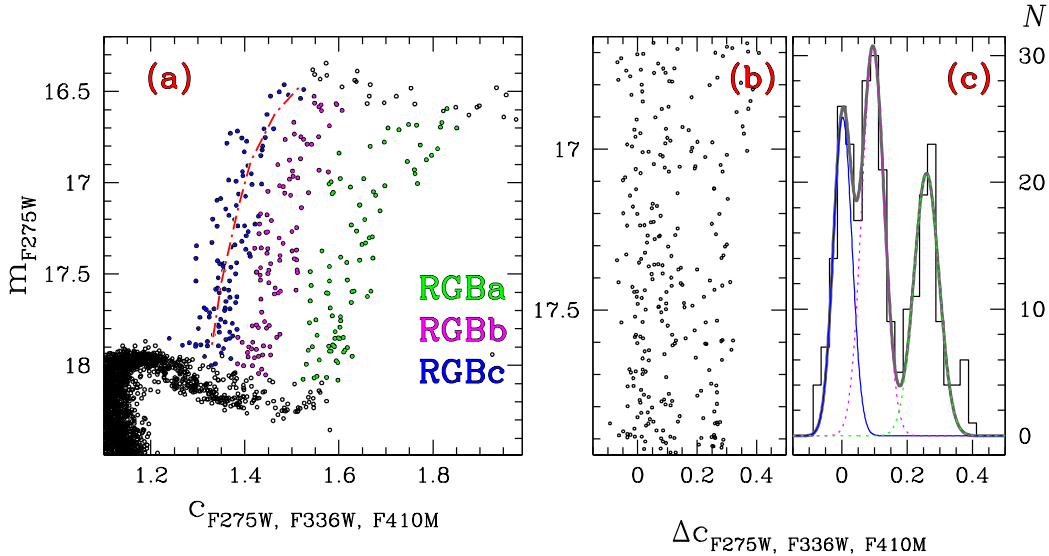}
%/home/milone/WORKS/SUMMARYNGC6752/MATCH336/FIGs/testrgb.macro go4
%/home/milone/WORKS/SUMMARYNGC6752/MATCH336/ZP4M/fig.macro go
   \caption{\textit{Panel (a):} Zoom in of the $m_{\rm F275W}$ versus
  $c_{\rm F275W, F336W, F410M}$ diagram of Fig.~\ref{275336390}c
  around the RGB. 
Selected RGBa, RGBb, and RGBc stars are plotted
  green, magenta, and blue respectively. \textit{Panel (b):}
  Verticalized diagram for RGB stars with $16.65<m_{\rm
    F275W}<17.95$. \textit{Panel (c):} Histogram of the distribution
  of stars shown in panel (b).
The gray line is the best-fitting least-square function defined as the sum of the green, the magenta, and the blue Gaussians.}
\label{selRGB}
\end{figure}

To extend our multi-wavelength study of the RGB, we attempted 
to identify the sequences corresponding to each stellar population along the RGB. We already noted that the $m_{\rm F275W}$ versus $c_{\rm F275W, F336W,
  F410M}$ diagram of Fig.~\ref{275336390} shows a triple RGB. In
Fig.~\ref{selRGB}, panel (a) we isolated by hand three groups of
stars. Hereafter, we will name these three groups as RGBa, RGBb, and
RGBc, colored green, magenta, and blue, respectively.  The red
fiducial line is obtained with a procedure similar to the one
introduced in Sect.~\ref{sec:MS} for MS stars. The only difference is
that for RGB stars we used a second-order polynomial to interpolate
the median color and magnitude values measured in the different
magnitude intervals.

The verticalized $m_{\rm F275W}$ versus $\Delta c_{\rm F275W, F336W, F410M}$ diagram 
is plotted in panel (b) of Fig.~\ref{selRGB}, while panel (c) of 
the same figure shows the histogram of the distribution in $\Delta c_{\rm F275W, F336W, F410M}$. The histogram is fitted by the sum of three Gaussians which 
we colored green, magenta, and blue respectively. From the area of the Gaussians we derive the relative fraction of RGBa, RGBb, RGBc stars
  to be (0.31$\pm$0.03, 0.41$\pm$0.02, 0.28$\pm$0.03).
%__________________________________________________________________
%__________________________________________________________________

\begin{figure}[ht!]
\centering
\epsscale{.75}
\plotone{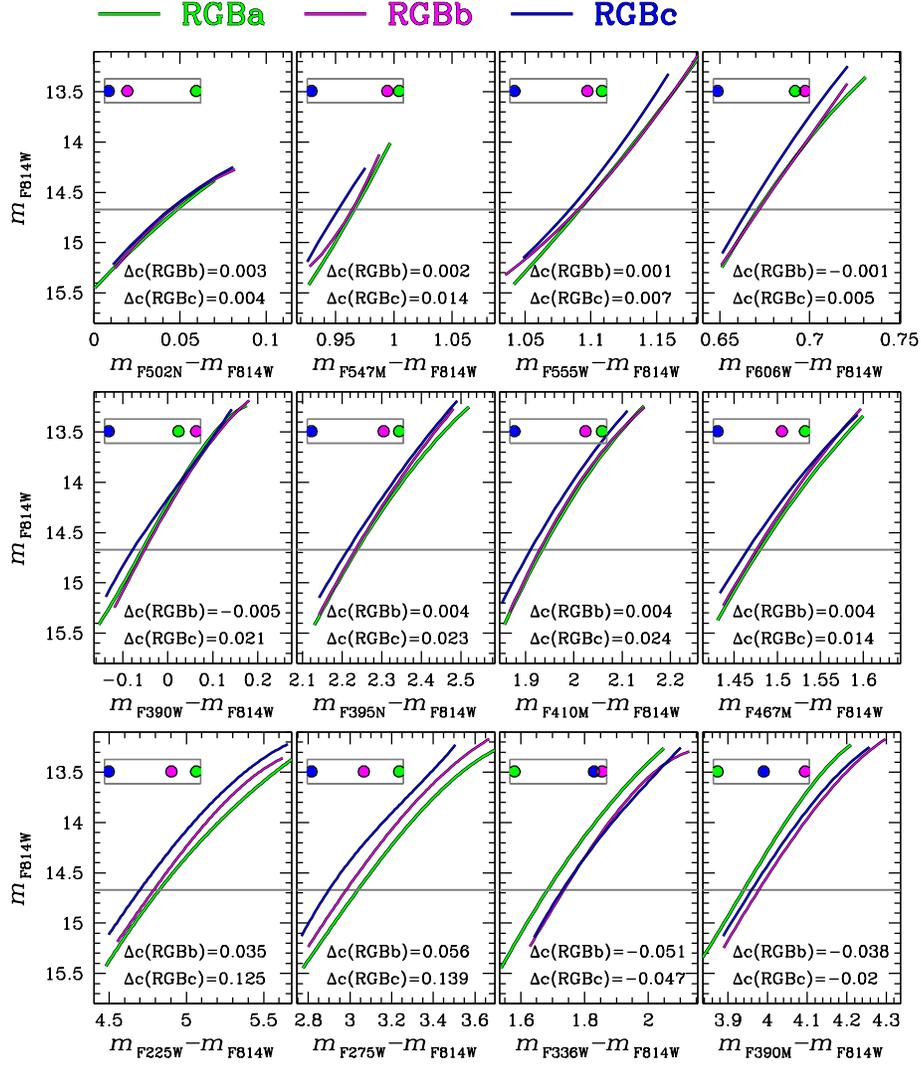}
%/home/milone/WORKS/SUMMARYNGC6752/MATCH336/FIGs/MSRL.macro go
   \caption{RGB fiducials in twelve $m_{\rm F814W}$ versus $m_{\rm
       X}-m_{\rm F814W}$ CMDs (X=F225W, F275W, F336W, F390M, F390W,
     F395N, F410M, F467M, F502N, F547M, F555W, and F606W). The color
     distance from the RGBa, measured at $m_{\rm
       F814W}^{\rm cut}=14.7$, is indicated in each panel.
The positions of RGBa, RGBb, and RGBc at $m_{\rm  F814W}^{\rm cut}$=14.7
are represented with green, magenta, and blue circles, respectively in the inset of each CMD.
 } 
\label{FIDUCIALsRGB}
\end{figure}

In order to get additional information on the three RGBs from all possible combinations
of magnitudes in the photometric passbands of our data set, 
we follow the same procedure for the RGB stars that we performed in Sect.~\ref{sec:MS} for the MS.
 Results are illustrated in Fig.~\ref{FIDUCIALsRGB}
where we show the fiducial polynomials  
in twelve $m_{\rm F814W}$ versus $m_{\rm
       X}-m_{\rm F814W}$ CMDs, where X=F225W, F275W, F336W, F390M, F390W,
     F395N, F410M, F467M, F502N, F547M, F555W, and F606W. In this analysis, we did not
     include F110W and F160W filters because, 
the RGB fiducials are poorly determined in these colors, on account of the small number of RGB stars that have IR photometry
This is consequence of three main facts: i) the IR/WFC3 camera has a smaller field of view than UVIS/WFC3, ii) due to its large pixel scale, the IR/WFC3 detector is more affected by crowding, hence the fraction of stars with high-accuracy photometry is smaller. iii) IR photometry saturates a couple of magnitudes above the MS turn-off.
%__________________________________________________________________
%__________________________________________________________________

 Both RGBb and RGBc are typically bluer than the RGBa, with the
 exception of  CMDs based on the $m_{\rm F336W}-m_{\rm F814W}$ and $m_{\rm
   F390W}-m_{\rm F814W}$ colors. In the other filters the color distance from
 the RGBa of both the RGBb and the RGBc increases with the color
 baseline, with the possible exception of $m_{\rm F390M}-m_{\rm
   F814W}$ and $m_{\rm  F395N}-m_{\rm F814W}$.
A comparison of Fig.~\ref{FIDUCIALs} and \ref{FIDUCIALsRGB} reveals
that the behaviour of the three RGBs and the three MSs 
is quite similar, and it will be discussed in  Sect.~\ref{sec:3msINT}.

\subsection{The chemical composition of the three stellar populations}
\label{chimica}
In the past three decades many spectroscopic studies have provided us with
 an accurate picture of the chemical composition of  NGC\,6752  
(e.g.\ Norris et al.\ 1981, Grundahl et al.\ 2002, Yong et
al.\ 2003, 2005, 2008, Carretta et al.\ 2007, 2012).  
We know that NGC\,6752 is a moderately metal poor
([Fe/H]$\sim -$1.6 Yong et al.\ 2005, Carretta et al.\ 2010) GC, with a large
star-to-star variations in O, N, Na, Mg, and Al. Nitrogen is
correlated with aluminum and sodium, and has a possible
small amplitude correlation with $\alpha$, Fe-peak, and $s$-process elements (Yong
et al.\ 2008). Both the Na-O and the Mg-Al anticorrelations have been
observed by Yong et al.\ (2005) and Carretta et al.\ (2007, 2009, 2012).
 
Str\"omgren photometry can provide additional, important information on the chemical
properties of the stellar populations. Grundahl et al.\ (2002) and Sbordone et al.\ (2011) have
demonstrated that the $c_{\rm y}$ index correlates with the nitrogen
abundance and hence can be used to identify different stellar
populations in GCs. Other combinations of Str\"omgren filters, 
are  sensitive to the chemical differences of GC stars (see, e.g.\, Grundahl et al.\ 2000, Marino et al.\ 2011, Carretta et al.\ 2011).     
%__________________________________________________________________
\begin{figure}[htp!]
\centering
\epsscale{.75}
\plotone{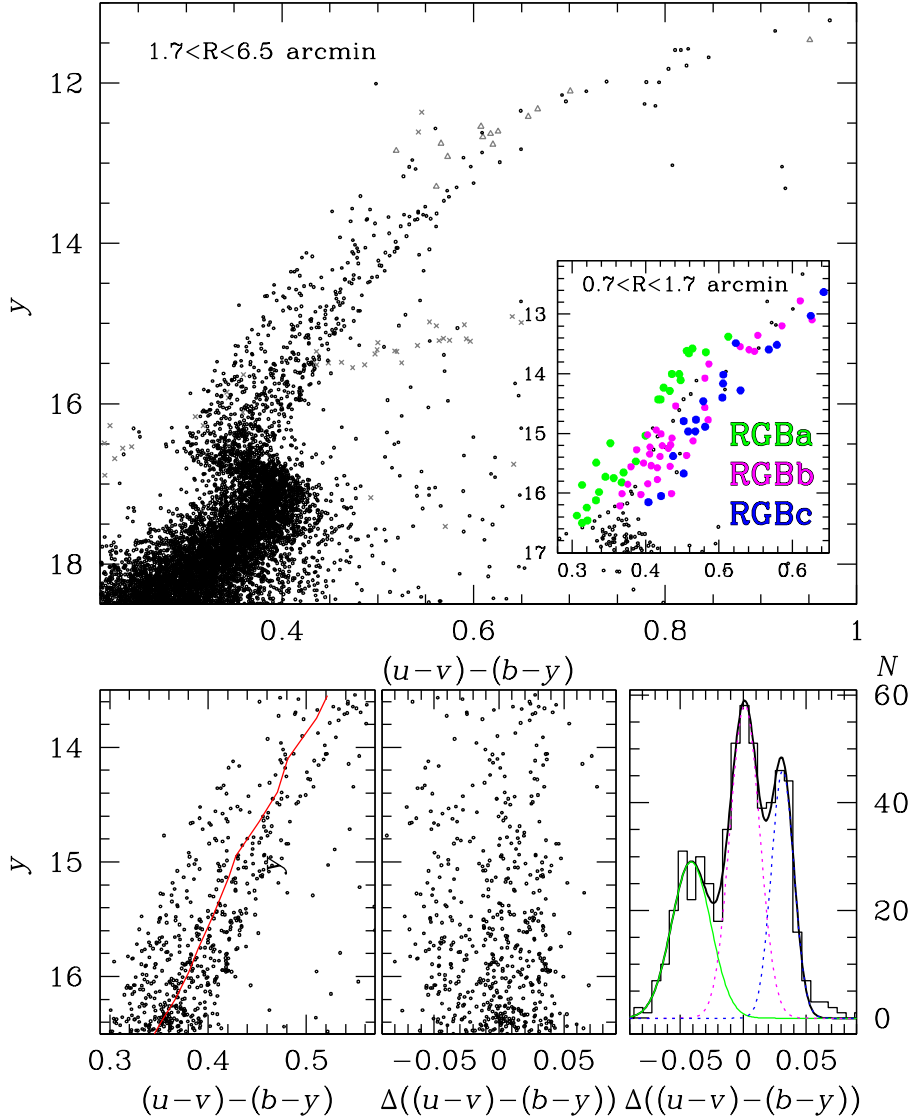}
%/home/milone/WORKS/SUMMARYNGC6752/GRUNDAHL/reddening/fig.macro
   \caption{\textit{Upper panel:} $y$ versus ($u-v$)$-$($b-y$)
     diagram for NGC\,6752 obtained using the Str\"omgren photometric
     catalog from Grundahl et al.\ (2002). Only stars with distance from the cluster center
     greater than 1.7 arcmin are plotted in this panel. 
    Asymptotic giant branch and HB stars, identified in the $b$ versus $v-y$ CMD are represented with gray triangles and crosses, respectively. 
     The inset is a zoom
   around the RGB for stars with distance from the center 0.7$<$R$<$1.7 arcmin where we colored green, magenta, and blue the RGBa,
   RGBb, and RGBc stars selected in Fig.~\ref{selRGB}. Lower panels
   illustrate the procedure to estimate the fraction of stars in each
   RGB. Left panel shows the fiducial line superimposed to the RGBb,
   the rectified CMD for RGB stars is plotted in the middle panel
   while right panel shows the histogram of the $\Delta$(($u-b$)$-$($v-y$))
   distribution with the best fit Gaussian colored green, magenta, and
   blue.}
\label{stromgren}
\end{figure}
%__________________________________________________________________

The $y$ versus ($u-b$)$-$($v-y$) diagram of NGC\,6752 is shown in
Fig.~\ref{stromgren}, upper panel (see Carretta et al.\ 2011 for a discussion on this color index). To avoid the central cluster regions, where the
photometric error is larger because of crowding, in the upper
panel of Fig.~\ref{stromgren} we show stars with a distance from the center 
larger than 1.7 arcmin. 
As already pointed out by Grundahl et al.\ (2002)
in Str\"omgren photometry CMDs the RGB shows three main components. We investigated whether these three RGBs correspond to the three populations we have identified in this paper. To do this, we cross-identified the stars in the  \textit{HST\/} and Str\"omgren catalog and
in the CMD in the inset of the upper panel of Fig.~\ref{stromgren} (which
includes only stars out to 1.7 arcmin from the cluster center), we colored 
green, magenta, and blue the RGBa, RGBb, and RGBc
stars identified in Fig.~\ref{selRGB}.   
Even if the stars common to both  \textit{HST\/} and Str\"omgren data sets 
are all located near the cluster center - and hence have larger photometric errors - 
it is clear that stars in RGBa, RGBb, and RGBc have different location (color) in the
CMD from Str\"omgren photometry.

In order to estimate the number of stars in each RGB, we have first isolated
the RGB region with $13.5<y<16.5$ and  distance from the center larger than 1.7
arcmin,   and drew by hand a fiducial line through the middle RGB as illustrated in the
 bottom-left panel of Fig.~\ref{stromgren}. Then, we verticalized the
 RGB following the procedure described in Sect.~\ref{sec:MS} (middle
 lower panel of Fig.~\ref{stromgren}). Finally, we
obtained the histogram of the $\Delta$(($u-b$)$-$($v-y$)) color distribution
 shown in the right panels of Fig.~\ref{stromgren}.    
 The histogram was least-squared fitted with three
 Gaussians.  
By summing the area under the three Gaussians, we find that the
fractions of RGBa, RGBb, and RGBc stars at radial distances between
 1.7 and 6.1 arcmin are 27$\pm$4\%, 44$\pm$4\%, and 29$\pm$3\%, respectively.
The quoted uncertainties are Poisson errors and represent a lower limit of the true errors.
All methods applied thus far to the MS and RGB, yield very similar
fractions of the $a$, $b$, and $c$ populations suggesting an
association between the MS and RGB for the three populations.

\begin{figure}[htp!]
\centering
\epsscale{.75}
\plotone{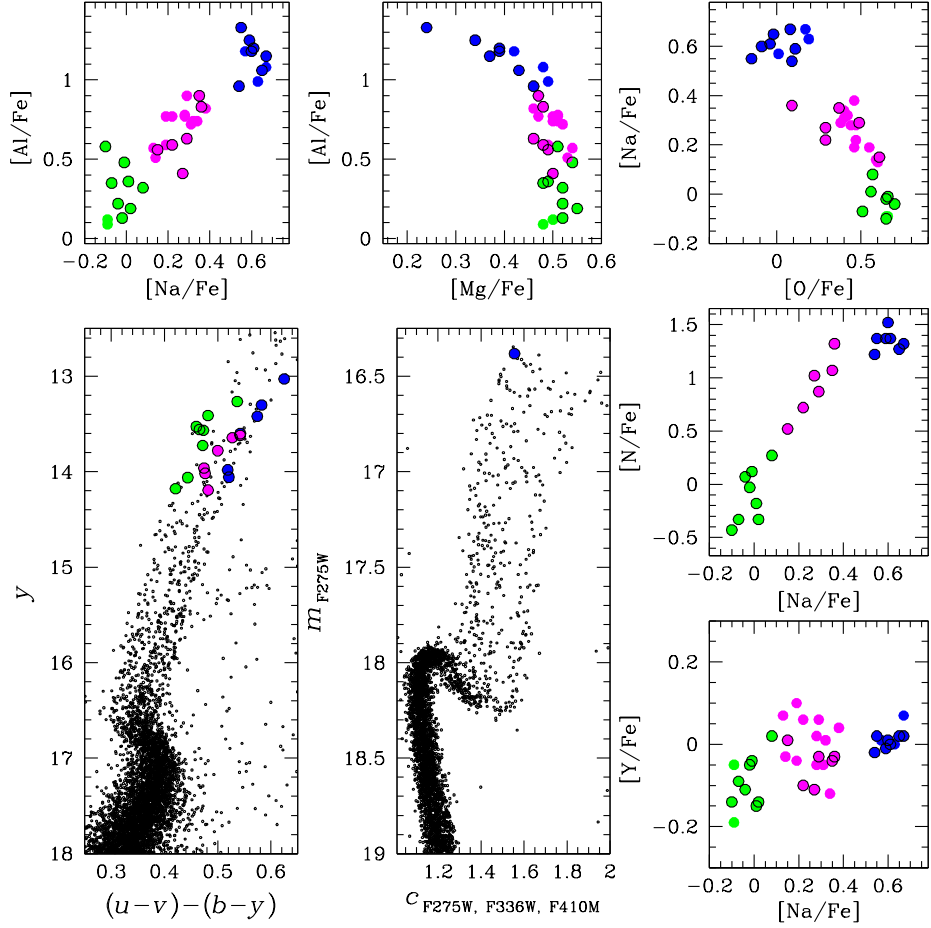}
%/home/milone/WORKS/SUMMARYNGC6752/GRUNDAHL/MATCHY/SPETTROSCOPIA/ide.macro go fig
   \caption{\textit{Upper} and \textit{right panels:} Correlations
     and anticorrelations among the abundances of several chemical
     species from Yong et al.\ (2003, 2008). We have selected three
     groups of stars with different [Na/Fe] and colored them green,
     magenta, and blue. In the bottom-left and central panel we mark
     the position of these stars in the $v$ versus ($u-v$)$-$($b-y$)
     diagram and the $m_{\rm F275W}$ versus $c_{\rm F275W, F336W, F410M}$ diagram.
Stars for which both spectroscopic and photometric measurements are available
are marked with black circles.}
\label{abundance}
\end{figure}

Available spectroscopic abundances allow us to better characterize
the chemical content of each single RGB.
Figure~\ref{abundance} reproduces the correlations among Al, Mg, N, Na, O 
using the chemical abundance measurements by Yong et al.\ (2005, 2008).
In the upper-right panel, we arbitrarily selected three groups 
of Na-poor, Na-intermediate, and Na-rich stars and colored them green,
magenta, and blue, respectively. These color codes are used
consistently in the other panels of this figure. 

As Yong and collaborators have already pointed out,
Na-rich stars are enhanced in Al,
N, and $s$-process elements, and depleted in O and
Mg. Na-intermediate stars are also Al, N-enhanced and O-depleted,
but their abundance variations  
are smaller, on average, than those of Na-rich stars. There is no significant
difference in the Mg and Y content of Na-poor and Na-intermediate
stars. It is interesting to note that the three groups of stars are
not chemically homogeneous, as they show star-to-star variation in the
abundance of some elements that are significantly larger than
observational errors (see Yong et al.\ 2008 for more details).

In the $y$ versus ($u-v$)-($b-y$) diagram of the left-bottom
panel of Fig.~\ref{abundance},
we mark with full dots the Na-poor, Na-intermediate,
and Na-rich stars selected in the upper-leftmost panel. The three RGBs
correspond to groups of stars with different light-elements
content, as first noticed by Grundahl et al.\ (2002). 
We have  \textit{HST\/} photometry for only one of the stars observed by
Yong et al.\ (2008). Its position in the 
$m_{\rm F275W}$ versus $c_{\rm F275W, F336W, F410M}$ diagram is shown in
the lower-middle panel and confirms that the RGBc is made of Na-rich
stars. 
%__________________________________________________________________
%__________________________________________________________________

The average abundance for 22 elements from Yong et al.\ (2008) for the three groups of RGB stars defined above are listed in Table~\ref{tab:abundance}.
We emphasize how the available spectroscopic measures offer us a precious opportunity
to characterize the chemical composition of stars of the three 
stellar populations in NGC\,6752.  

%----------------------------------------------------------------
\begin{table}[htp!]
\begin{center}  
\scriptsize {
\caption{Average chemical abundance for the three RGB components.}
\begin{tabular}{|l|ccc|ccc|ccc|}
\hline
\hline
  population & & Population a & & & Population b & & & Population c & \\
\hline
  Element & abundance [dex]& $\sigma$ & N & abundance [dex]& $\sigma$ & N  & abundance [dex]& $\sigma$ & N \\
\hline
\hline
 {\rm [O/Fe] } &    0.65$\pm$0.04 & 0.11 &  10  &      0.43$\pm$0.03 & 0.13 &   18  &  0.03$\pm$0.04 & 0.11 &   10 \\
 {\rm [N/Fe] } & $-$0.11$\pm$0.09 & 0.25 &   8  &      0.92$\pm$0.13 & 0.28 &    6  &  1.35$\pm$0.04 & 0.10 &    7 \\
 {\rm [Na/Fe]} & $-$0.03$\pm$0.02 & 0.06 &  10  &      0.26$\pm$0.02 & 0.08 &   18  &  0.61$\pm$0.02 & 0.05 &   10 \\
 {\rm [Mg/Fe]} &    0.51$\pm$0.01 & 0.02 &  10  &      0.49$\pm$0.01 & 0.02 &   18  &  0.40$\pm$0.02 & 0.07 &   10 \\
 {\rm [Al/Fe]} &    0.28$\pm$0.05 & 0.16 &  10  &      0.70$\pm$0.03 & 0.14 &   18  &  1.14$\pm$0.04 & 0.12 &   10 \\
 {\rm [Si/Fe]} &    0.27$\pm$0.02 & 0.07 &  10  &      0.33$\pm$0.01 & 0.05 &   18  &  0.35$\pm$0.01 & 0.04 &   10 \\
 {\rm [Ca/Fe]} &    0.21$\pm$0.03 & 0.08 &  10  &      0.24$\pm$0.02 & 0.09 &   18  &  0.27$\pm$0.02 & 0.06 &   10 \\
 {\rm [Sc/Fe]} & $-$0.05$\pm$0.01 & 0.04 &  10  &   $-$0.04$\pm$0.01 & 0.04 &   18 &$-$0.04$\pm$0.01 & 0.04 &   10 \\
 {\rm [Ti/Fe]} &    0.10$\pm$0.02 & 0.06 &  10  &      0.14$\pm$0.01 & 0.04 &   18  &  0.15$\pm$0.01 & 0.03 &   10 \\
 {\rm [V/Fe] } & $-$0.34$\pm$0.05 & 0.16 &  10  &   $-$0.29$\pm$0.03 & 0.12 &   18 &$-$0.25$\pm$0.03 & 0.08 &   10 \\
 {\rm [Mn/Fe]} & $-$0.50$\pm$0.04 & 0.11 &  10  &   $-$0.44$\pm$0.01 & 0.06 &   18 &$-$0.45$\pm$0.01 & 0.04 &   10 \\
 {\rm [Fe/H]}  & $-$1.65$\pm$0.02 & 0.07 &  10  &   $-$1.61$\pm$0.01 & 0.02 &   18 &$-$1.61$\pm$0.01 & 0.01 &   10 \\
 {\rm [Co/Fe]} & $-$0.03$\pm$0.03 & 0.09 &  10  &   $-$0.00$\pm$0.02 & 0.06 &   18 &$-$0.06$\pm$0.02 & 0.07 &   10 \\
 {\rm [Ni/Fe]} & $-$0.06$\pm$0.02 & 0.07 &  10  &   $-$0.06$\pm$0.01 & 0.04 &   18 &$-$0.03$\pm$0.01 & 0.03 &   10 \\
 {\rm [Cu/Fe]} & $-$0.66$\pm$0.03 & 0.09 &  10  &   $-$0.59$\pm$0.01 & 0.05 &   18 &$-$0.60$\pm$0.01 & 0.04 &   10 \\
 {\rm [Y/Fe] } & $-$0.09$\pm$0.02 & 0.06 &  10  &   $-$0.01$\pm$0.02 & 0.06 &   18  &  0.01$\pm$0.01 & 0.02 &   10 \\
 {\rm [Zr/Fe]} &    0.07$\pm$0.05 & 0.15 &  10  &      0.20$\pm$0.02 & 0.07 &   18  &  0.21$\pm$0.03 & 0.09 &   10 \\
 {\rm [Ba/Fe]} & $-$0.09$\pm$0.04 & 0.13 &  10  &   $-$0.12$\pm$0.03 & 0.13 &   16  &  0.05$\pm$0.02 & 0.07 &    9 \\
 {\rm [La/Fe]} &    0.12$\pm$0.02 & 0.02 &   2  &      0.10$\pm$0.01 & 0.04 &   12  &  0.13$\pm$0.04 & 0.06 &    3 \\
 {\rm [Ce/Fe]} &    0.28$\pm$0.03 & 0.09 &  10  &      0.25$\pm$0.01 & 0.04 &   18  &  0.28$\pm$0.02 & 0.06 &   10 \\
 {\rm [Nd/Fe]} &    0.23$\pm$0.01 & 0.04 &  10  &      0.22$\pm$0.01 & 0.05 &   18  &  0.23$\pm$0.01 & 0.04 &   10 \\
 {\rm [Eu/Fe]} &    0.31$\pm$0.03 & 0.10 &  10  &      0.30$\pm$0.02 & 0.08 &   18  &  0.34$\pm$0.03 & 0.10 &   10 \\
\hline \hline
\label{tab:abundance}
\end{tabular}
}
\end{center}
\end{table}
\subsection{Some considerations on the spectroscopic and photometric observational evidence of multiple stellar populations}

The Na-O anticorrelation has been proposed as a possible proof of the
presence of multiple stellar generations in star clusters. 
Several authors suggested that the different populations 
of stars in GCs
can be identified on the basis of their position along the Na-O abundance plane, with the first generation including stars with oxygen and sodium abundance similar to halo field stars at the same metallicity, with
remaining Na-rich/O-poor to be considered as part of a
second generation  (e.g.\, Kraft et al.\ 1994). 

Spectroscopic studies show that the Na-O anticorrelation is a common property among  GCs (e.g.\, Ramirez \& Cohen 2002, Carretta et al.\ 2009, 2010 and references therein).
However, there clearly is a problem we have not solved yet. 
Figure~12 from the compilation by  Ramirez \& Cohen (2002) and Fig.~1 by Carretta et al.\ (2010) shows an almost continuous  distribution of stars in the Na versus O plane, despite the fact that photometric evidence in many of the clusters included in that figure (e.g.\, NGC\,2808, NGC\,6397, NGC\,6752 as shown
in the present paper) show multimodal, maybe discrete, sequences in the CMD when high-accuracy photometry on images collected with the appropriate filters is used\footnote{Note however that, in some cases, multimodal distribution in Na and O have been detected also from high-resolution spectroscopic (see e.g.\, Yong et al.\ 2008, Marino et al.\ 2008, Lind et al.\ 2011 for the cases of NGC\,6752, M\,4, and NGC\,6397).}.
Why do we have this possible difference between the spectroscopic and the
photometric manifestation  of the multiple stellar populations in GCs?

\begin{figure}[htp!]
\centering
\epsscale{.45}
%/home/milone/WORKS/SUMMARYNGC6752/GRUNDAHL/MATCHY/SPETTROSCOPIA/ide.macro
\plotone{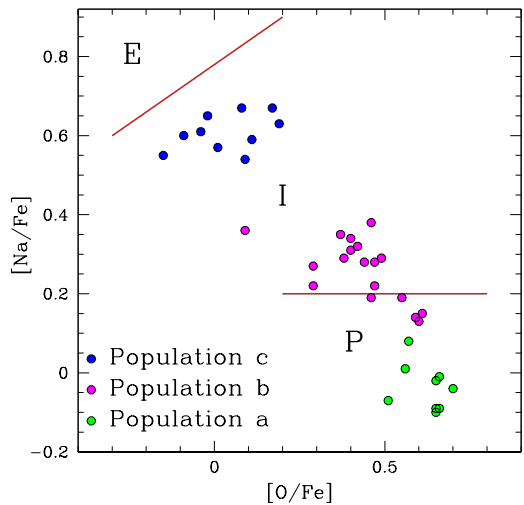}
   \caption{Reproduction of the Na-O anticorrelation from Yong et al.\ (2008). 
Stars belonging to the populations a, b, and c defined in this paper are colored green, magenta, and blue, respectively. The red segments indicate the separation for the P, I, and E components suggested by Carretta et al.\ (2009).
}
\label{fig:app}
\end{figure}

Carretta et al.\ (2009) suggested criteria to separate stellar populations on the basis of their position in the Na-O plane.
They defined as primordial (P) component all stars with [Na/Fe] ratio in the range between $\rm [Na/Fe]_{\rm min}$ and $\rm [Na/Fe]_{\rm min}+$0.3 where $\rm [Na/Fe]_{\rm min}$ is the minimum value of the ratio [Na/Fe] ratio estimated by eye. 
The remaining stars are considered all second-population stars, and have been further divided into two groups. Stars with ratio [O/Na]$>-$0.9 dex belong to the intermediate (I) population, while those with [O/Na]$<-$0.9 dex are defined as 
extreme (E) population. Clearly, this separation is arbitrary and not based on any
feature (gaps or peak) in the NaO diagram,  and it has no clear physical meaning.
As an example, in Fig.~\ref{fig:app}, we apply the same criteria as proposed by Carretta et al.\ (2009) to the sample of NGC\,6752 stars studied by Yong at al.\ (2008) and further analyzed in Sect.~\ref{chimica}.
The two red segments define the regions in the Na-O plane populated by P, I, and E stars, and are determined following the recipes by Carretta and collaborators. The stars of the populations a, b, and c identified in this paper are colored green, magenta, and blue, respectively.
Figure~\ref{fig:app} shows that:
i) the Carretta et al.\ (2009) P component includes all the stars in population a but is contaminated by  population b stars; ii) the I component contains both population b and population c stars, and 
(iii) no stars belong to the E component. Apparently, the general criteria introduced by Carretta et al.\ (2009) do not apply to NGC\,6752.

The main problem here is not related to the meaning of the Carretta
et al.\ (2009) definition. 
The problem is that there still is an inconsistency between what
we observe with spectroscopy (a continuous distribution along the Na vs. O plane) and what accurate, high precision photometry tells us, i.e.\ that
most clusters host distinct, separate evolutionary branches in the CMD. It is important to understand whether this is just the consequence of the (internal)
errors in the measurement of Na and O abundances, or whether there is some underlying physical reason we have not understood, yet.
%

%__________________________________________________________________
%__________________________________________________________________

\section{The Helium abundance of the three stellar populations of NGC\,6752}
\label{sec:3msINT}
In this section, we will use the multicolor  \textit{HST\/} photometry to further characterize the chemistry of the different stellar populations in NGC\,6752
and estimate the helium abundances of its three stellar populations.
 The multi-dimensional space of our CMDs makes it difficult to visualize fitting isochrones to our sequences.  To make the comparison easier, we will quantify the color separation between the sequences at 
two fiducial points at the level of the MS and the RGB.

\begin{figure}[htp!]
\centering
\epsscale{.45}
\plotone{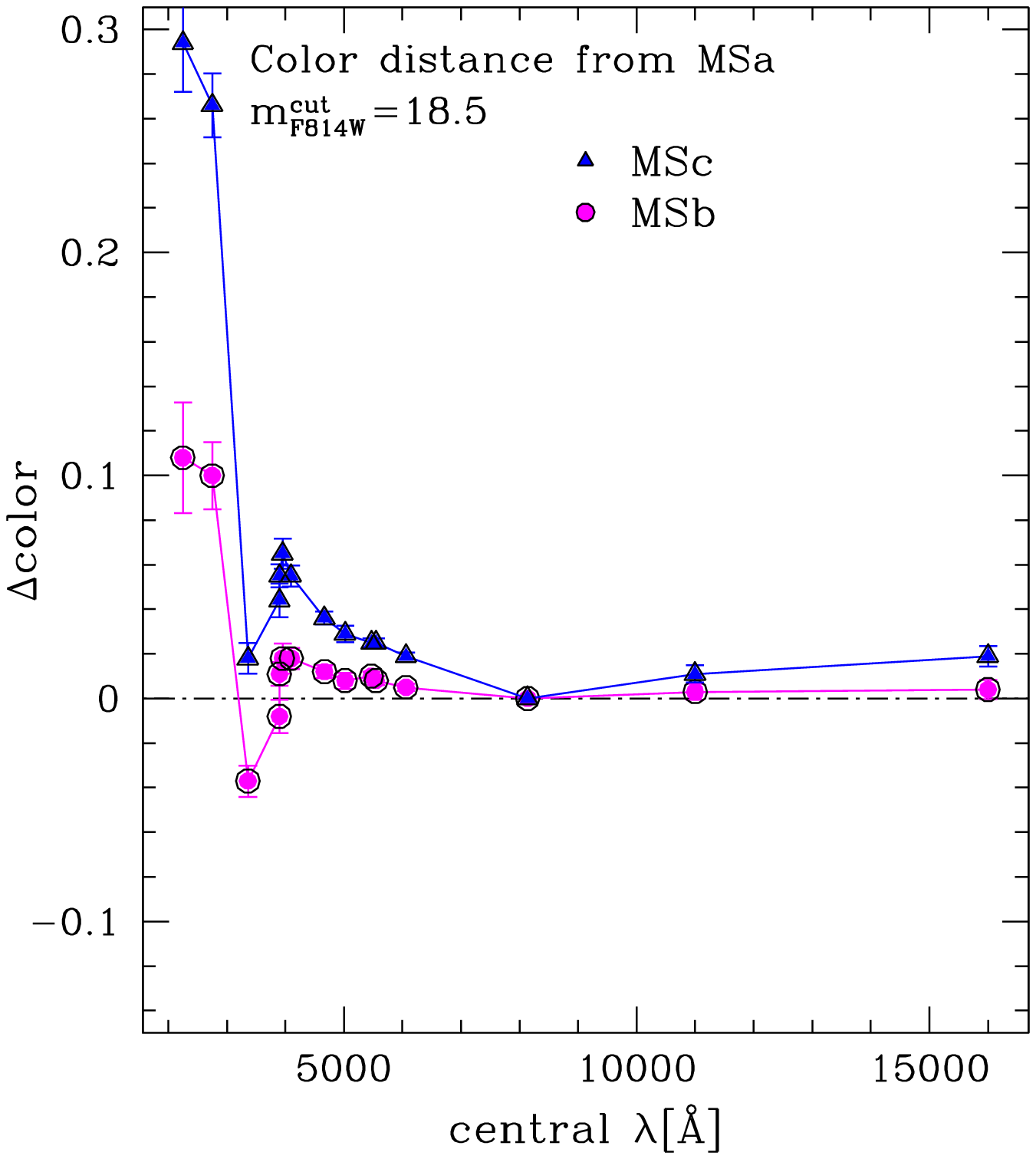}
\plotone{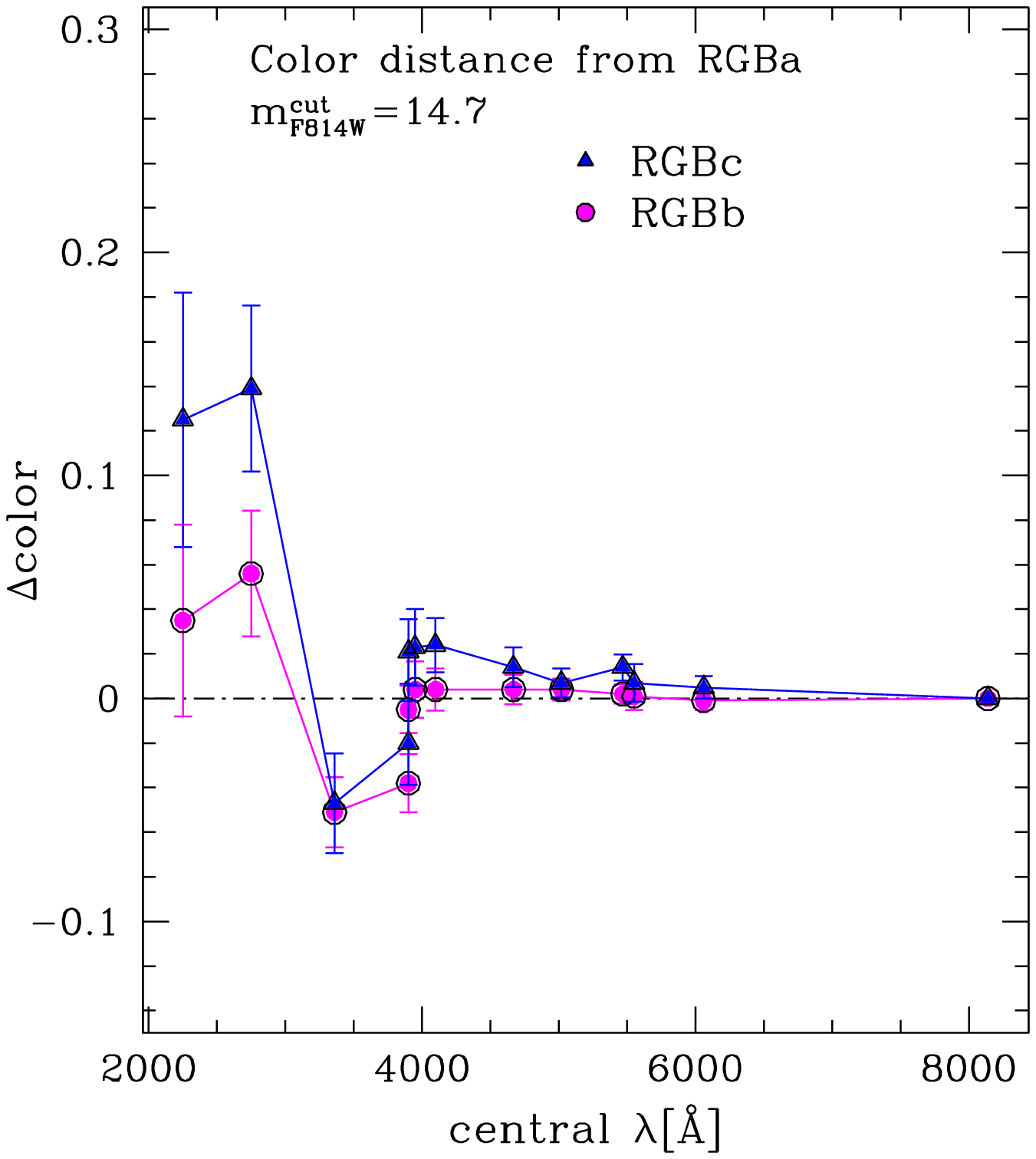}
%/home/milone/WORKS/SUMMARYNGC6752/MATCH814/FIGs/MSRL.macro go2 go3
%/home/milone/WORKS/SUMMARYNGC6752/MODELLI/testRGBnew.macro go
   \caption{
     \textit{Left Panel:} $m_{\rm X}-m_{\rm F814W}$ (or
     $m_{\rm F814W}-m_{\rm X}$) color 
     distance between MSb and MSa (magenta circles) and
     between MSc and MSa (blue triangles) as a function of the
     central wavelength of the  X filter. 
     \textit{Right Panel:} 
     Color distance between RGBb and RGBa (magenta circles) and between RGBc and the RGBa (blue triangles). The color distances of MS and RGB sequences are measured at the reference magnitude $m_{\rm F814W}^{\rm cut}$=18.5 and $m_{\rm F814W}^{\rm cut}$=14.7, respectively.
 }
\label{MSdis}
\end{figure}

In all fourteen CMDs of Fig.~\ref{FIDUCIALs}, 
we calculated the color differences between MSa, MSb, and MSc at the
reference magnitude $m_{\rm    F814W}^{\rm cut}$=18.5. 
 These color differences are calculated by subtracting from the color of the MSa fiducial at $m_{\rm    F814W}^{\rm cut}$=18.5 the color of the MSb (or MSc) 
fiducial at the same luminosity.
Left panel of Fig.~\ref{MSdis} shows
 the measured color difference as a function of the
 central wavelength of the $m_{\rm X}$ filter.
 
%__________________________________________________________________
%__________________________________________________________________
%
It is now clear why the $c_{\rm F275W, F336W, F410M}$ index is an efficient tool to  identify multiple sequences in the CMD. Figure~\ref{MSdis} indeed shows that both the $m_{\rm F275W}-m_{\rm F336W}$ and the $m_{\rm F336W}-m_{\rm F410M}$ color provide large separations between the MSa (or the RGBa) and the other two MSs (RGBs).
 The MSa (RGBa) is redder than the MSb and the MSc (RGBb and RGBc) in $m_{\rm F275W}-m_{\rm F336W}$, but it moves to the blue in $m_{\rm F336W}-m_{\rm F410M}$. 
Therefore, the color index $c_{\rm F275W, F336W, F410M}$=$(m_{\rm F275W}-m_{\rm F336W})-(m_{\rm F336W}-m_{\rm F410M})$ maximizes the separation among the sequences.\footnote{By following a similar approach we suggest that other color indices $c_{\rm F336W, F410M, F814W}$=$(m_{\rm F336W}-m_{\rm F410M})-(m_{\rm F410M}-m_{\rm F814W})$ can be  powerful tools to detect multiple stellar populations in the CMD of GCs.
The later is less efficient than $c_{\rm F275W, F336W, F410M}$, but it uses a wavelength range that is accessible to ground-based telescopes.}

In order to compare these observations with expectations from synthetic photometry, we followed a procedure 
that has already been used in previous publications
(Milone et al.\ 2012a,b, Bellini et al.\ 2013,
see also Sbordone et al.\ 2011). Briefly, we used the BaSTI isochrones (Pietrinferni et al.\ 2004, 2009) for 
the populations listed in Table~\ref{tab:parameters}, and determined $T_{\rm eff}$ and $\log g$ for a MS star with $m_{\rm F814W}^{\rm cut}$=18.5.
To do this, we assumed $E(B-V)=0.03$ 
and  $(m-M)_{\rm V}=13.19$ in agreement with the values of reddening and distance modulus listed in the Harris (1996, updated as in December, 2010). 
 For ACS/WFC filters we used the extinction coefficients tabulated 
by Bedin et al.\ (2005) for a 
cold (T=4000K) star. For the WFC3/UVIS filters, we linearly interpolated the coefficients tabulated by Bedin et al.\ (2005), and for WFC3/IR we adopted the values listed in the York Extinction Solver website\footnote{http://www3.cadc-ccda.hia-iha.nrc-cnr.gc.ca/community}.
For each of the three populations, we adopted the abundances of the 
22 elements listed in Table~\ref{tab:abundance}, which represent the average abundance of the three stellar populations.
Since carbon abundance was not measured by Yong et al.\ (2008) we adopted the [C/Fe] values from Carretta et al.\ (2005) 
and take [C/Fe]=0.15 for population a, and [C/Fe]=$-$0.15 for both population b, and c.

We assumed that the population a has primordial helium content (Y=0.246),
and adopted for both the population b, and c different helium abundances
 as described in the following.
We used the ATLAS12 code (Kurucz 2005, Castelli 2005, Sbordone 2007)
to calculate model atmospheres by using the specific chemical
composition of each stellar population, and assuming the temperatures
and gravities listed in Table~\ref{tab:parameters}. Then, we used the
SYNTHE code (Sbordone et al.\ 2007) to synthesize the spectrum from
1000\AA\ to 20\,000\AA\, and the resulting spectra were convolved with the
transmission curves of the system formed by the telescope, the camera, and each of our filters to produce the synthetic magnitudes and colors for each photometric band.  

\begin{figure}[htp!]
\centering
\epsscale{1.05}
\plotone{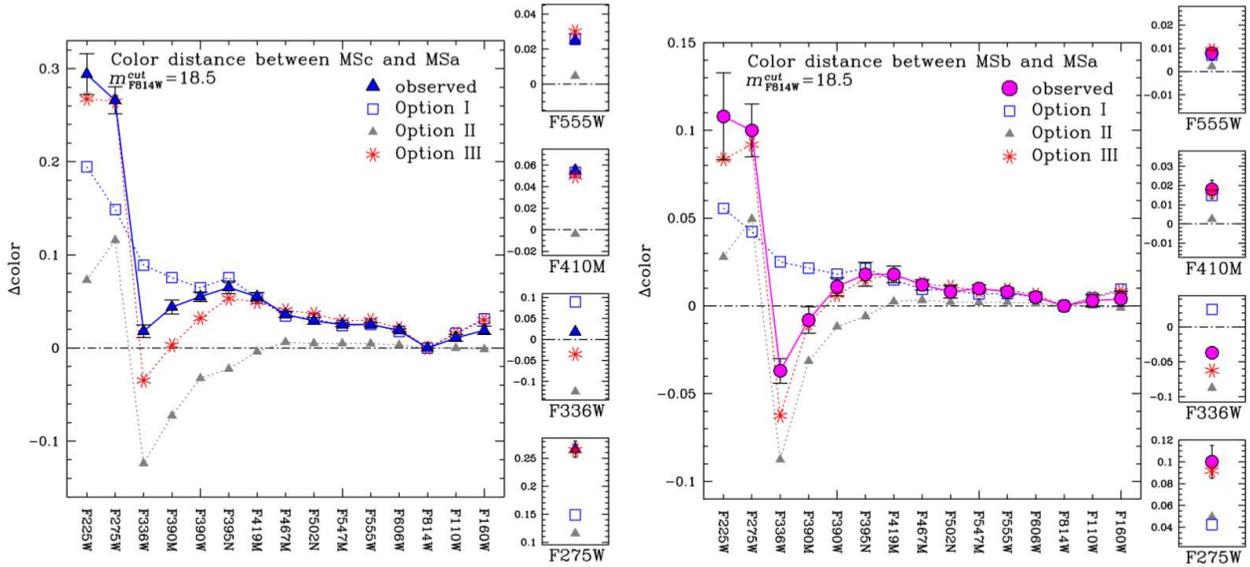}
%/home/milone/WORKS/SUMMARYNGC6752/MODELLI/
   \caption{
Color separations at different color baselines for
the ridge lines of the MSc and the MSa (left), 
 and the MSb and the MSa (right) at $m_{\rm F814W}=$18.5.
Observations are plotted as blue triangles and magenta circles for the MSc and the MSb, respectively, while the color differences
expected from theoretical Options I, II, and III are shown as blue
squares, gray triangles, and red asterisks.   
On the right of each panel, 
the small boxes correspond to regions centered on the F275W, F336W, F410M, and F555W bands.
}
\label{modMS}
\end{figure}

The adopted chemical compositions for the three different stellar
populations a,b,c identified within NGC\,6752 , (MSs, SGBs, and RGBs),
are tabulated in Table~\ref{tab:parameters} for three different
options.
In all these three options, population a is assumed to have a
canonical helium abundance and the chemical composition in Table~2. 
Instead, the chemical compositions and the helium fraction (Y) of populations b and c,  are different for different options.
In details: 
\begin{itemize}
\item  In Option~I, the three populations have the same chemical composition, 
       but three different values for the helium content (Y$=$0.246, 0.254, and 0.275). 
\item  In Option~II, the three populations have the same helium fraction, 
       but the three chemical compositions given in Table~2.
\item  In Option~III, the three populations have the three chemical composition given in Table~2, 
       \textit{and} three different values of helium as described below. 
 We estimated
for population a, b, and c, Y$=$0.246, 0.254, and 0.275, respectively. 
\end{itemize}
The comparison of the observed $m_{\rm X}-m_{\rm F814W}$ differences between 
MSa and MSc against the synthetic ones are shown in  the left panel of Fig.~\ref{modMS}. 

The blue squares  indicate the color differences corresponding to Option~I, where we assumed for  
the two stellar populations the same element abundance as for MSa, 
but a different helium content. 
Similar to what observed in the cases of 47\,Tuc and NGC\,6397, we find a good agreement in most bands but a significant disagreement for filters bluer than $\sim$4000\AA, and conclude that helium cannot be 
the only parameter responsible for the color differences between MSa and MSc. 
This confirms that the observed light-element abundance of the three stellar populations  play a fundamental role in the MS morphology. 

In Option~II we assumed for MSc stars the same He content as for the MSa, and the chemical composition listed in Table~\ref{tab:parameters}. The colors resulting from this option are represented by gray triangles in Fig.~\ref{modMS}, and indicate a significant disagreement between observed and synthetic colors.
It is worth noting that the observed light-element difference among the three stellar populations have a negligible effect for filters redder than $\sim$4000\AA as already shown by Sbordone et al.\ (2011).

Finally, in Option~III (red asterisks)
we assumed differences in both helium and chemical abundances. 
To determine the value of Y that best reproduces the observed points, we generated a grid of synthetic spectra by assuming different helium abundances, with Y ranging from 0.246 to 0.290 in steps of $\Delta$Y=0.001. 
For each synthetic spectrum we determined the $m_{\rm X}-m_{\rm F814W}$ color distance between MSc and MSa corresponding to each X filter: ($\Delta{\rm color}_{\rm syn}$), and calculated
 $d(Y)$=$|\Delta{\rm color}_{\rm syn}-\Delta{\rm color}|$
 where $\Delta{\rm color}$ is the observed color distance between the two MSs in that filter.

The helium abundance ($Y_{*}$) that minimizes $d(Y)$ is assumed as the best estimate of Y for the filter X. The helium content of MSc is then calculated as the weighted mean of the available $Y_{*}$ measurements by using only those filters redder than F395N we obtain Y=0.273$\pm$0.002.  
The quoted error comes from the weighted mean, and does not take into account neither the uncertainties of the synthetic spectra, nor possible errors on the value we assumed for the primordial helium.
Note that we have excluded from this analysis all the UV and far-UV filters as they are very sensitive to small variations of light-element abundance.
Option III (red asterisks), provides the best agreement with observations though not completely satisfactory at short wavelengths.

The right panel shows the comparison of the observed $m_{\rm X}-m_{\rm F814W}$ color differences between the MSa and the MSb against the synthetic one. Again in this case the best fit is given by the Option III, 
and the agreement between synthetic and observed photometry is
 much better than what we obtained for the MSc and MSa color differences.
By using the procedure described above we obtain for MSb Y=0.253$\pm$0.001.

\begin{figure*}[htp!]
\centering
\epsscale{.65}
\plotone{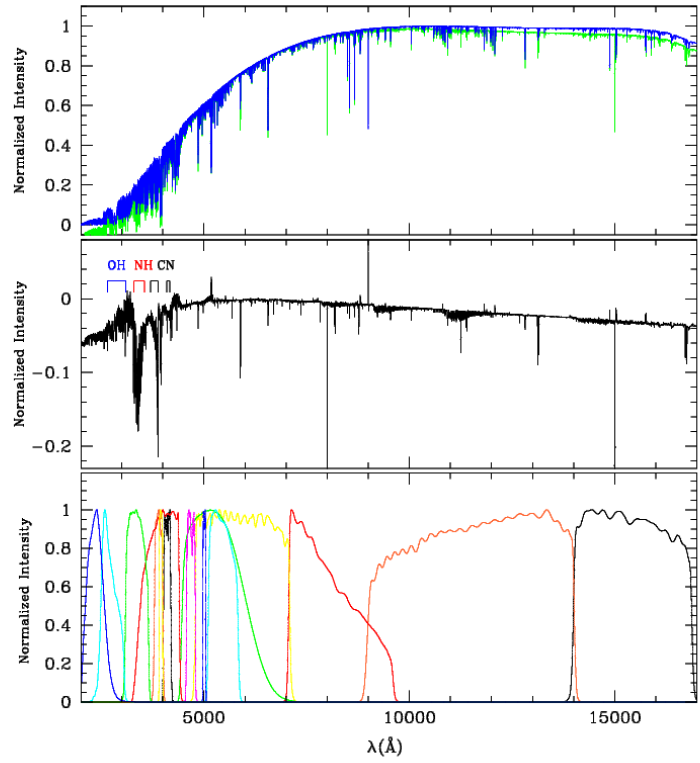}
%/home/milone/WORKS/SUMMARYNGC6752/MODELLI/testMSau.macro go3
   \caption{
\textit{Upper Panel}: comparison of the synthetic spectra of a
MSa star (green) and a MSc stars (blue). \textit{Middle Panel}: Difference between the spectra of the MSa and the MSc star.
 The location of relevant molecular features are indicated.
\textit{Lower Panel}: Normalized-to-peak response of the \textit{HST\/} filters used in this paper. }
\label{spettri}
\end{figure*}

The synthetic spectra of a MSa and a MSc star, as calculated for the Option III, are shown in the upper panel of Fig.~\ref{spettri}. The difference between the two spectra is in the middle panel, while the lower panel shows the (normalized-to-peak) band-passes of the filters used in this paper.

In summary, the observed color differences between the three MSs are consistent with three populations with different helium and light-element abundances. 
Specifically the MSa corresponds to the first stellar population with primordial He and O-rich/N-poor stars, the MSc is made of stars enhanced in He and N, but depleted in O, and MSb stars has intermediate He and C, N, O abundances.   

%__________________________________________________________________
%__________________________________________________________________

%__________________________________________________________________
%__________________________________________________________________

\begin{figure*}[htp!]
\centering
\epsscale{1.05}
\plotone{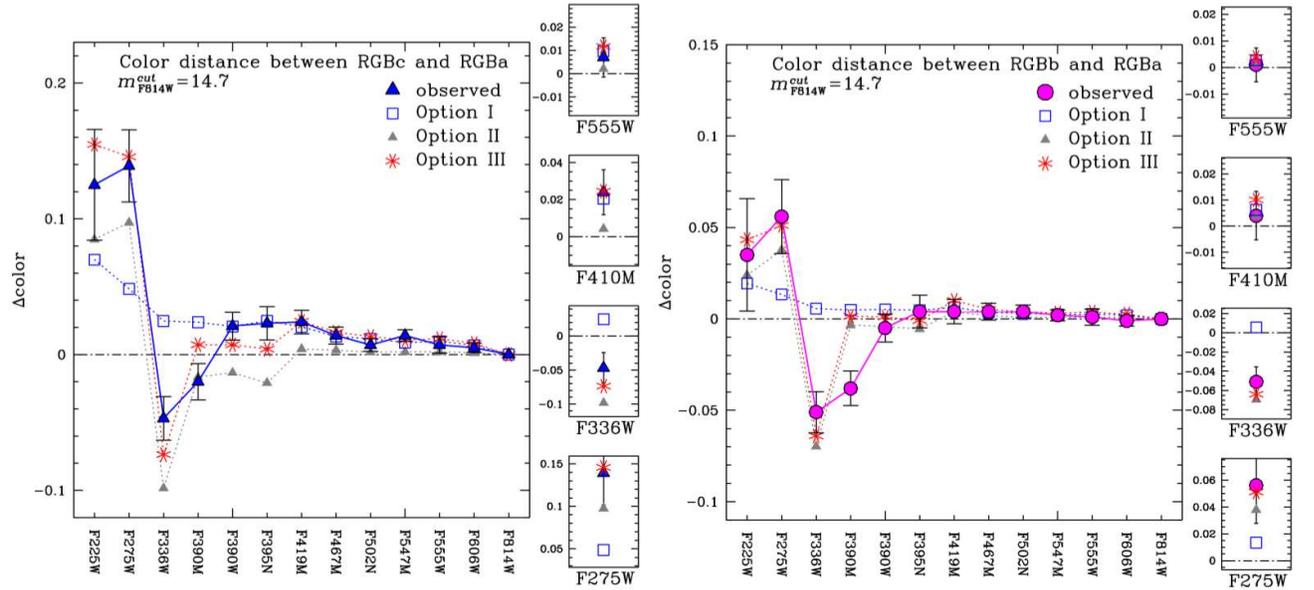}
%/home/milone/WORKS/SUMMARYNGC6752/MODELLI/testRGBnew go go2 mr testMS go2b
   \caption{
\textit{Left Panel:} $m_{\rm X}-m_{\rm F814W}$  color 
     separation between RGBc and RGBa as a function of the
     central wavelength of the  corresponding X filter. \textit{Right Panel:} Color separation
     between RGBb and RGBa. The color distances are measured at
     the reference magnitude $m_{\rm F814W}^{\rm cut}$=14.7.
Observations are plotted as 
blue triangles for RGc and magenta circle for RGBa,
while the color differences
expected from theoretical Options I, II, and III are shown as blue
squares, gray triangles, and red asterisks.
The small boxes correspond to regions centered on the F275W, F336W, F410M, and F555W bands.
}
\label{modRGB}
\end{figure*}

In Fig.~\ref{modRGB}, for the three RGBs  we repeated the same analysis as done for
the MSs.  
We have calculated the color distance of RGBb and RGBc from RGBa 
at $m_{\rm F814W}^{\rm cut}=$14.7. In Fig.~\ref{modRGB}, these color
differences are plotted as a function 
of each filter and labeled with its corresponding effective wavelength. 

Observations are compared with synthetic colors calculated for a RGB
stars with $m_{\rm F814W}^{\rm cut}$=14.7 by using the same procedure
described for MS stars, and for the same three Options I, II,
and III listed in Table~\ref{tab:parameters}. 
Similar to what found for the MSs, the color differences between the
three RGBs can be best reproduced by Option III, which reinforces the idea
previously proposed that RGBa, RGBb,
and RGBc stars are the progeny of MSa, MSb, and MSc stars, respectively.  
Specifically, by using the procedure described above for the MSs we obtain for RGBc and RGBb helium abundances of Y=0.272$\pm$0.005, and Y=0.255$\pm$0.004, 
respectively.
%__________________________________________________________________
%__________________________________________________________________

\begin{table}[htp!]
\begin{center}
\scriptsize {
\caption{Parameters used to simulate synthetic spectra of an MSa, MSb, and
an MSc star with $m_{\rm F814W}$=18.5, and an RGB star with $m_{\rm
  F814W}$=14.7 for the three assumed options. For all the
populations we assumed [Fe/H] = $-$1.6. The adopted chemical
composition is given in Table~\ref{tab:abundance}.}

\begin{tabular}{ccccc}
\hline
\hline
MS (Option) & $T_{\rm eff}$ & $\log g$ & $Y$ & chemical composition (see Table~\ref{tab:abundance}) \\
\hline
MSa (all)   & 5445 & 4.68 & 0.246 & as for population a \\
MSb (I)     & 5470 & 4.68 & 0.254 & as for population a \\
MSb (II)    & 5445 & 4.68 & 0.246 & as for population b \\
MSb (III)   & 5470 & 4.68 & 0.254 & as for population b \\
MSc (I)     & 5534 & 4.69 & 0.275 & as for population a \\
MSc (II)    & 5445 & 4.68 & 0.246 & as for population c \\
MSc (III)   & 5534 & 4.69 & 0.275 & as for population c \\
\hline                                     
RGBa (all)  & 5343 & 3.26 & 0.246 & as for population a \\
RGBb (I)    & 5351 & 3.26 & 0.254 & as for population a \\
RGBb (II)   & 5343 & 3.26 & 0.246 & as for population b \\
RGBb (III)  & 5351 & 3.26 & 0.254 & as for population b \\
RGBc (I)    & 5373 & 3.25 & 0.275 & as for population a \\
RGBc (II)   & 5343 & 3.26 & 0.246 & as for population c \\
RGBc (III)  & 5373 & 3.25 & 0.275 & as for population c \\
\hline
\hline
\label{tab:parameters}
\end{tabular}
}
\end{center}
\end{table}
%----------------------------------------------------------------

\subsection{The radial distribution of the three stellar populations}
\label{sec:rd}
\begin{figure}[htp!]
\centering
\epsscale{.55}
\plotone{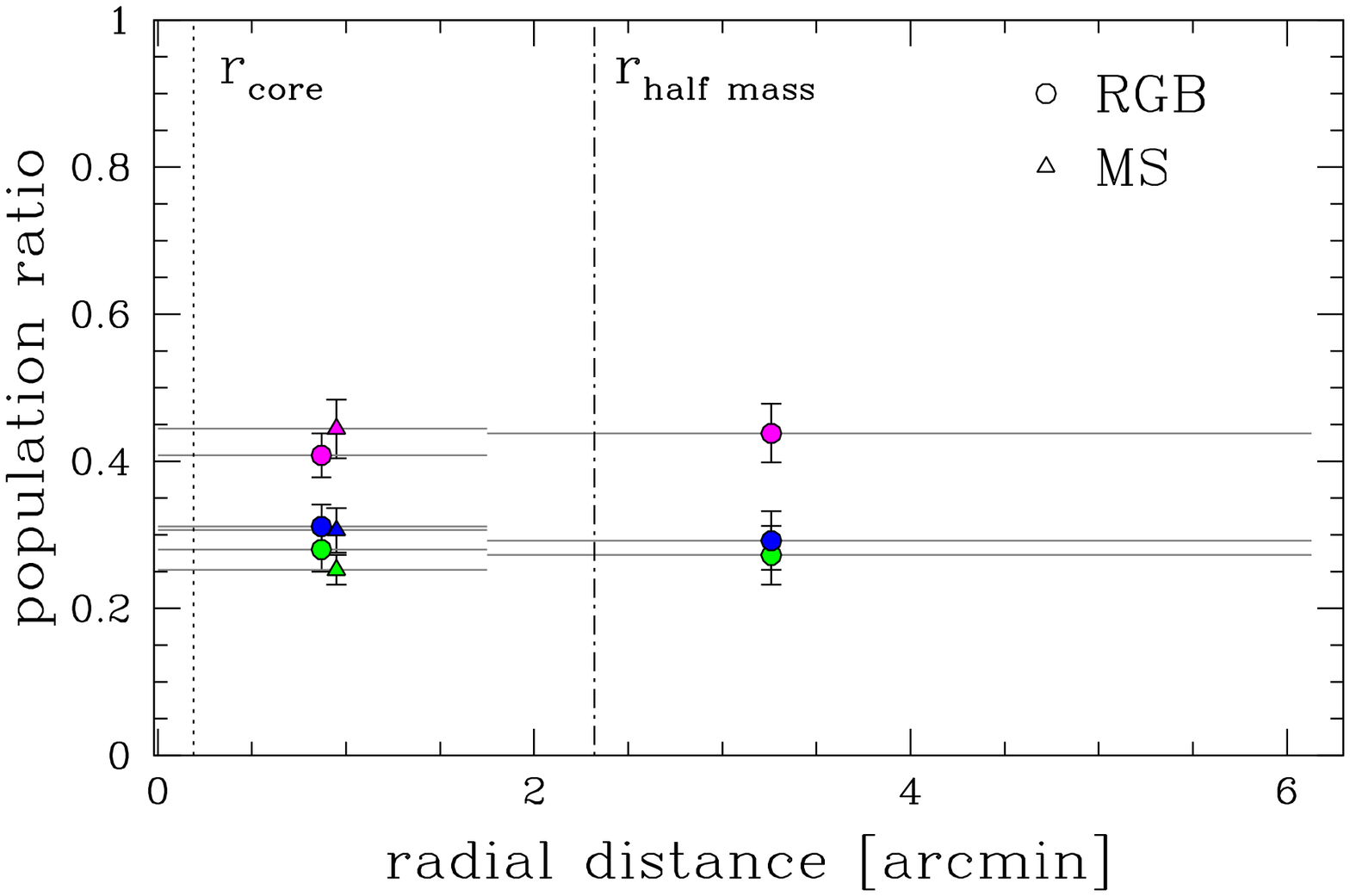}
\plotone{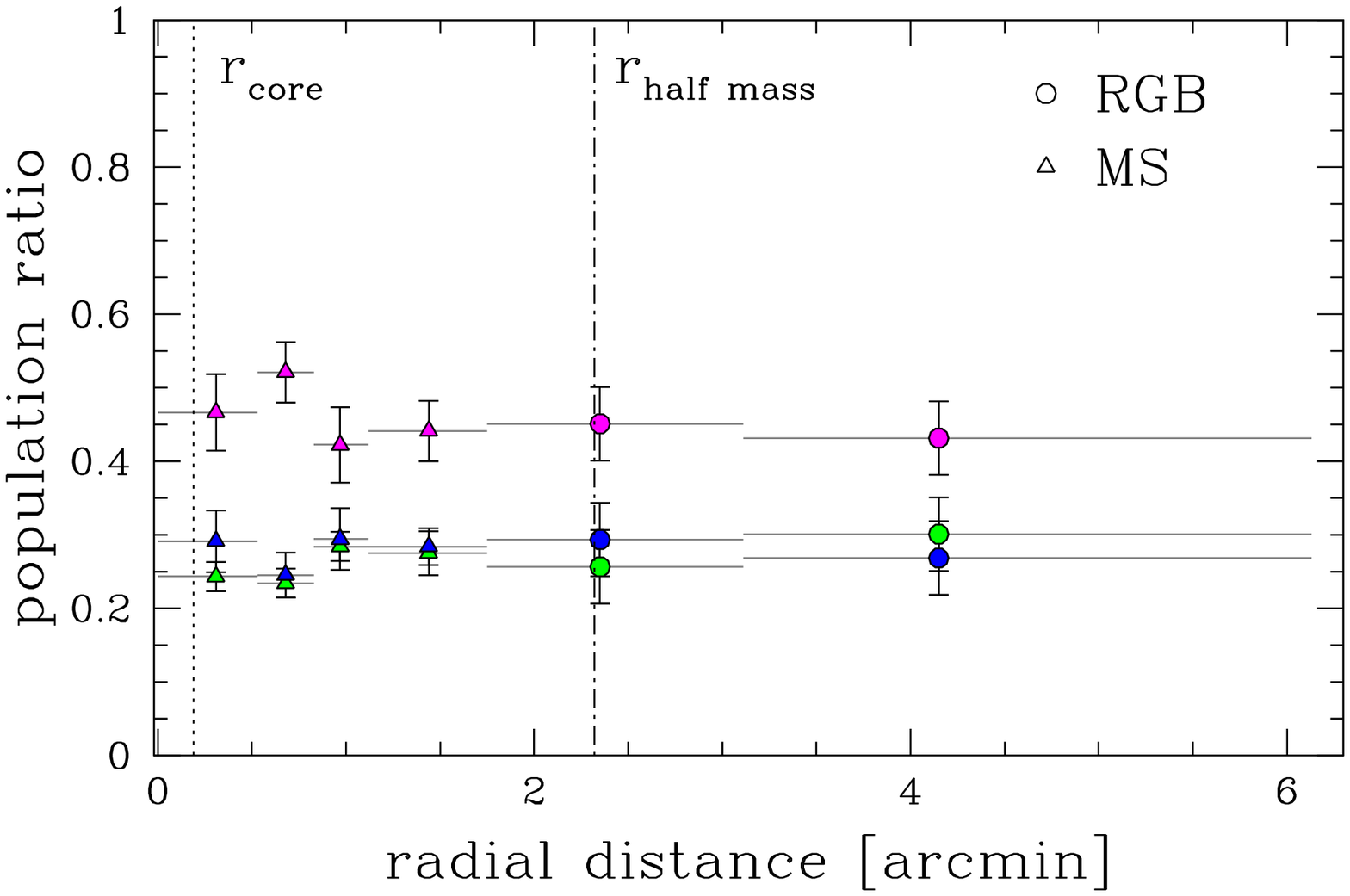}
%/home/milone/WORKS/SUMMARYNGC6752/MATCH336/FIGs/RD.macro go
   \caption{Radial distribution of the fraction of population a (green
     symbols), population b (magenta symbols), and population c (blue
     symbols) with respect to the total number of (population a +
     population b + population c) of stars. Circles and triangles
     refer to the measures obtained from RGB and MS stars,
     respectively. The dotted and the dashed-dotted vertical lines mark the 
     core and the half-mass radius, respectively.
      In the upper panel  we have used 
one single radial interval for {\it HST} and one for ground-based data. In the lower panel we have divided the {\it HST} field of view into four radial bins, and the region with radial distance larger than 1.7 arcmin 
(ground-based data)
into two bins.  See text for details.}
\label{RD}
\end{figure}

In order to calculate the radial distribution of the three stellar populations
in NGC\,6752, we  first divided our sample into two regions. 
The first one includes stars with radial distance from the cluster center smaller than 1.7 arcmin,
and is covered by \textit{HST\/} observations. The second one contains stars 
 with radial distances from the center between 1.7 and 6.5 arcmin, which are included 
in the ground-based catalog.
  The upper panel of Fig.~\ref{RD} shows 
 the fractions of MSa, MSb, and MSc stars with respect to the total 
 number of MS stars as green, magenta, and blue triangles, and the
 fractions of RGBa, RGBb, and RGBc stars with respect to all RGB population
 as green, magenta, and blue circles. 
\footnote{ We note here that the fraction of stars in the different MSs has been estimated by comparing stars in small magnitude intervals (see Sect.~3). Since these stars have similar luminosities and the color differences between MSa, MSb, and MSc stars is quite small, incompleteness is not an issue in this comparison. In any case, the completeness level is greater than $0.67$ for the adopted magnitude interval, with $m_{\rm F814W}<20.3$.}

 Then, we further divided the region with radial distance smaller than 1.7 arcmin into four circular subregions, each 
one containing almost the same number of MS stars. For each region, we calculated the fraction of MSa, MSb, and MSc stars by using the procedure already described in Sect.~\ref{sec:MS}.
 Similarly, we divided the region with radial distance between 1.7 and 6.5 arcmin into two parts, containing almost the same number of RGB stars and calculated the fraction of RGBa, RGBb, and RGBc stars as in Sect.~\ref{chimica}. Results are shown in the lower panel of Fig.~\ref{RD}.
Note that, due to the relatively small number of RGB stars in the {\it HST} field we preferred to limit the analysis based on {\it HST} data to MS stars.

The fraction of population a, population b, and population c stars are listed in Table~4, where we also included the minimum and the maximum radius of the circular region ($R_{\rm min}$ and $R_{\rm max}$), and the average 
radial distance  of the MS or RGB stars used to estimate the populations ratio ($R_{\rm ave}$). The later is calculated as the mean radius of the MS or RGB stars in that bin.

 There is no significant radial trend in the relative numbers of the 
 three stellar populations within six arcmin from the cluster center. 
  Apparently these results do not support the conclusions by Kravtsov et al.\ (2011) who, analysed ground-based images, and reported a strong difference in the radial distribution between the RGB subpopulations. 
 These authors found that the change in the fraction in the two RGBs occurs at radial distance close to the half-mass radius of the cluster ($r_{\rm h}$=2.34 arcmin, Harris 1996, 2003) and becomes much stronger at larger radial radii. An extension of the analysis presented in this paper to larger radial distances from the cluster center is mandatory to properly characterize
the radial distributions of the different subpopulations.

 As pointed out by the referee, the relaxation time of NGC\,6752 at the time the secondary generations formed is a fundamental ingredient for a proper interpretation of these results.  The estimate of this quantity  is beyond the objectives of this paper and is complicated by the fact that NGC 6752 (as, in general, all GCs showing evidence of multiple stellar populations)
may have been significantly more massive at the time of the formation of their stellar generations.
(see e.g.\ Conroy et al.\ 2012, Goudfrooij et al.\ 2011, D'Ercole et al.\ 2010). The populations ratio listed in Table~4 at several radial distance from the cluster center can provide useful contraints for the models of formation and evolution of stellar populations in GCs.

%__________________________________________________________________
%__________________________________________________________________

\begin{table}[!htp]
\begin{center}
\scriptsize {
\begin{tabular}{ccccccc}
\hline
\hline
 $R_{\rm min}$ & $R_{\rm max}$ & $R_{\rm ave}$ &  & population ratio& & sequence \\ 
\hline
 & & & a & b & c &     \\
\hline
0.00 & 1.70 & 0.95 & 0.25$\pm$0.02 &  0.44$\pm$0.04 &  0.31$\pm$0.03 &  MS  \\
0.00 & 1.70 & 0.87 & 0.28$\pm$0.03 &  0.41$\pm$0.03 &  0.31$\pm$0.03 &  RGB \\
1.70 & 6.13 & 3.26 & 0.27$\pm$0.04 &  0.44$\pm$0.04 &  0.29$\pm$0.04 &  RGB \\
\hline                                                                                 
0.00 & 0.53 & 0.31 & 0.24$\pm$0.02 &  0.47$\pm$0.05 &  0.29$\pm$0.042&  MS  \\
0.53 & 0.83 & 0.68 & 0.23$\pm$0.02 &  0.52$\pm$0.04 &  0.25$\pm$0.031&  MS  \\
0.83 & 1.12 & 0.97 & 0.28$\pm$0.02 &  0.42$\pm$0.05 &  0.29$\pm$0.042&  MS  \\
1.12 & 2.33 & 1.44 & 0.28$\pm$0.03 &  0.44$\pm$0.04 &  0.28$\pm$0.025&  MS  \\
1.70 & 3.11 & 2.35 & 0.26$\pm$0.05 &  0.45$\pm$0.05 &  0.29$\pm$0.05 &  RGB \\
3.11 & 6.13 & 4.15 & 0.30$\pm$0.05 &  0.43$\pm$0.05 &  0.27$\pm$0.05 &  RGB \\     
\hline
\hline
\label{tab:RD}
\end{tabular}
}
\caption{ Fraction of population a, population b, and population c stars 
 calculated in different circular
regions with different radial distance from the 
cluster center. The minimum and maximum radius ($R_{\rm min}$ and $R_{\rm max}$) 
of each region are listed together with the average radial distance of stars in each region ($R_{\rm ave}$).
The last column indicates the sequence of the CMD (MS or RGB) used to estimate the populations ratio.}
\end{center}
\end{table}
%----------------------------------------------------------------

\section{Summary}
\label{sec:discussion}
\begin{figure}[htp!]
\centering
\epsscale{.45}
%/home/milone/WORKS/SUMMARYNGC6752/MATCH336/FIGs/CMD.macro go6c go6d go7c go7d
\plotone{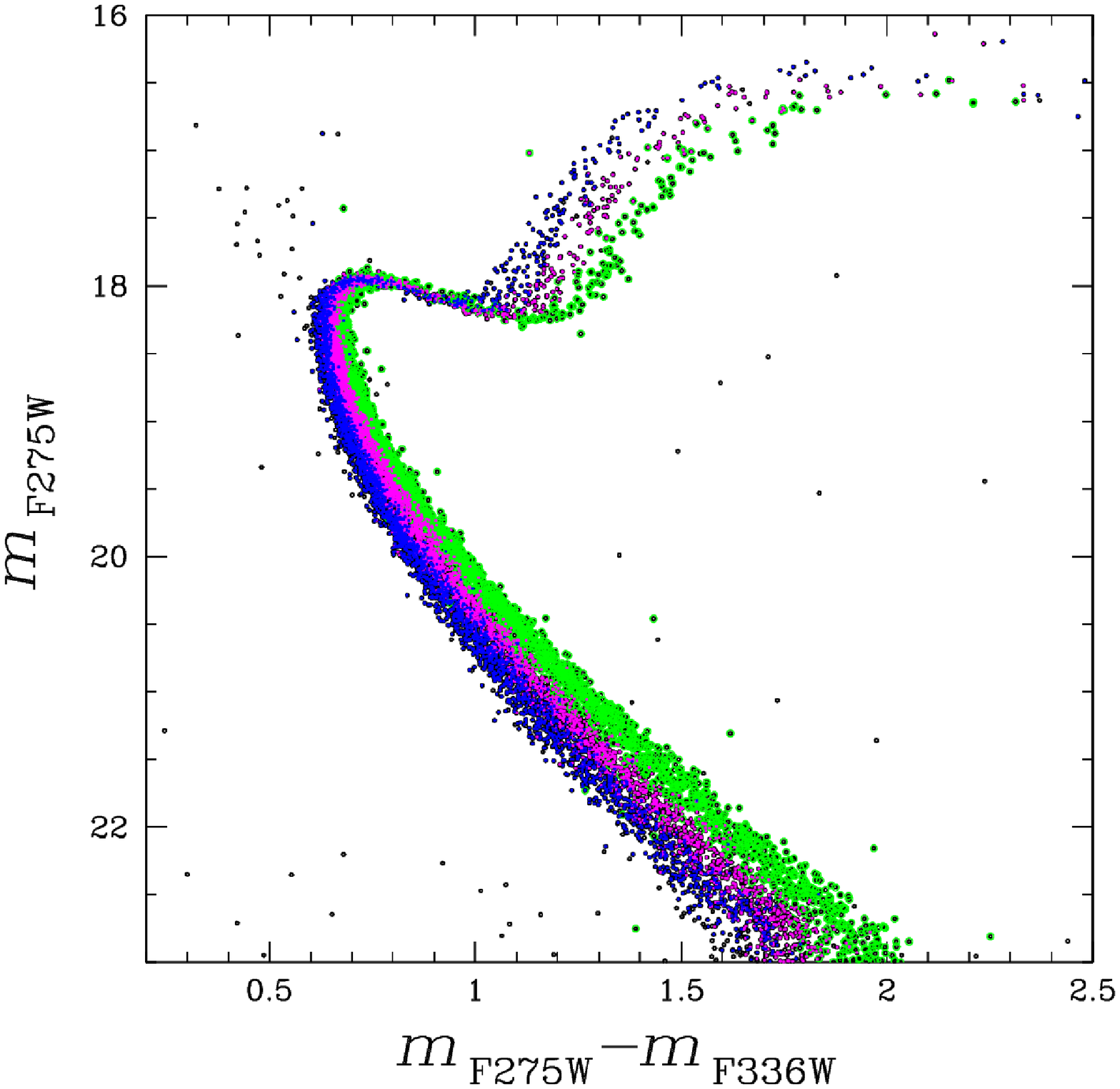}
\plotone{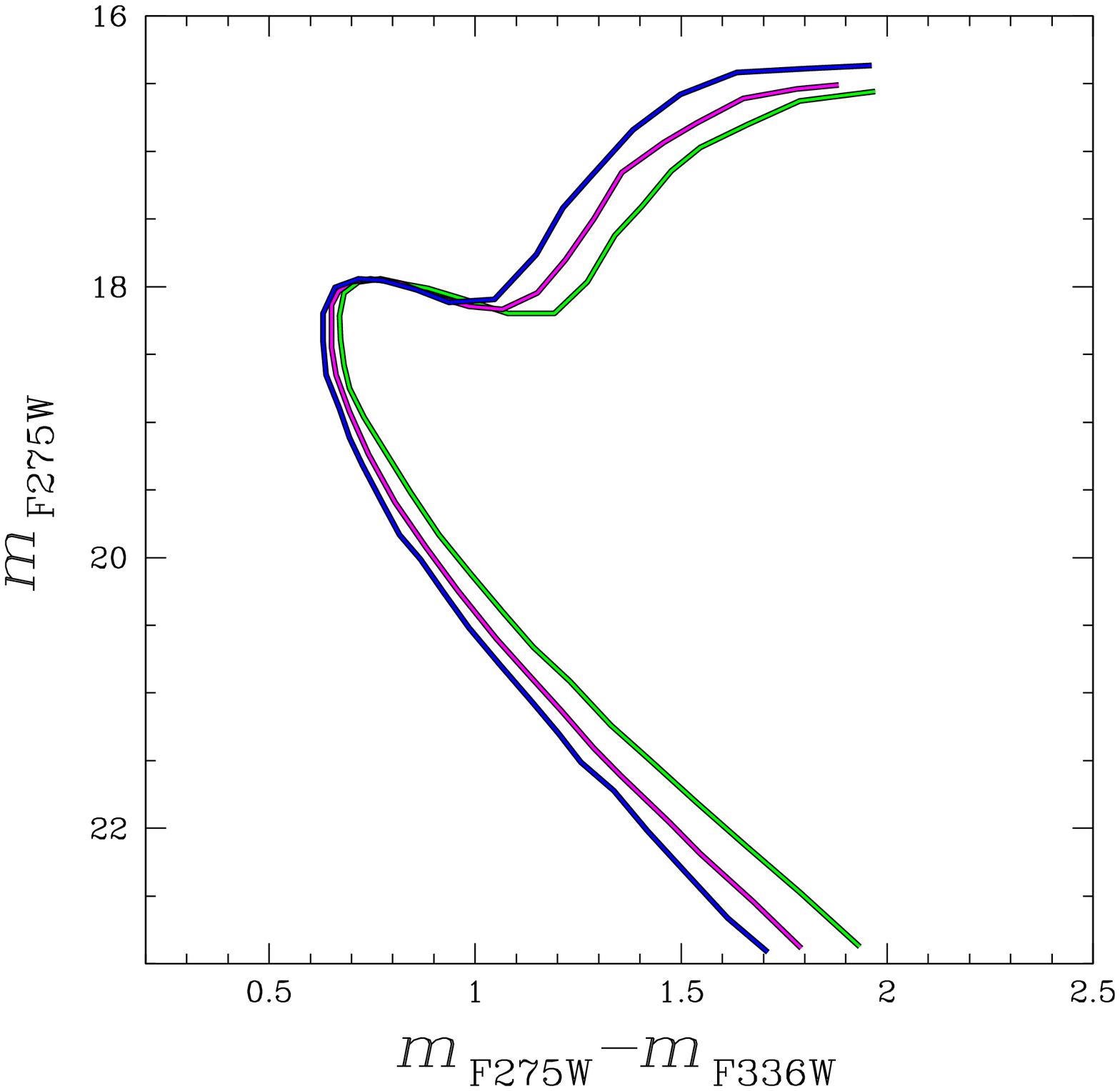}
\plotone{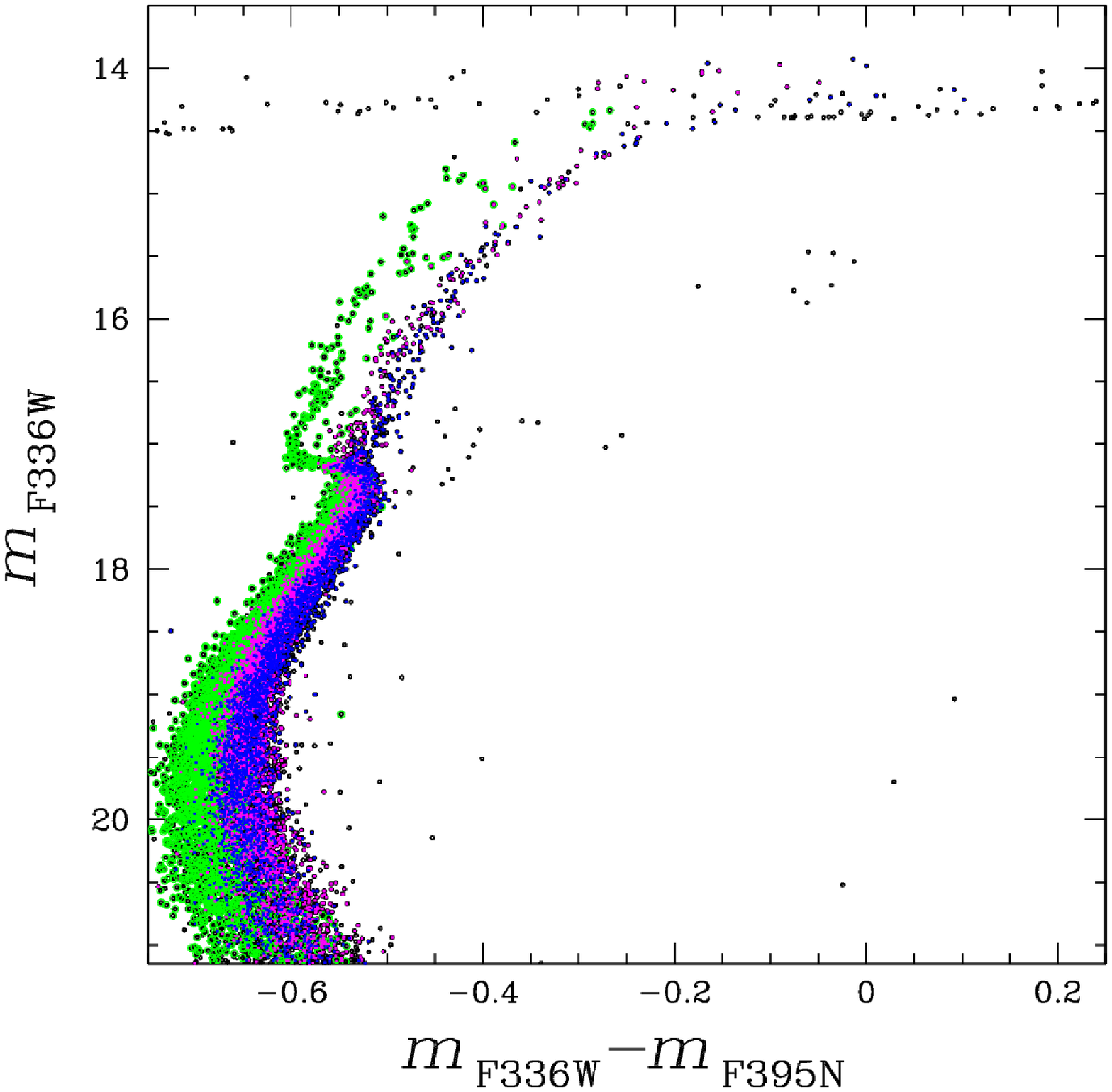}
\plotone{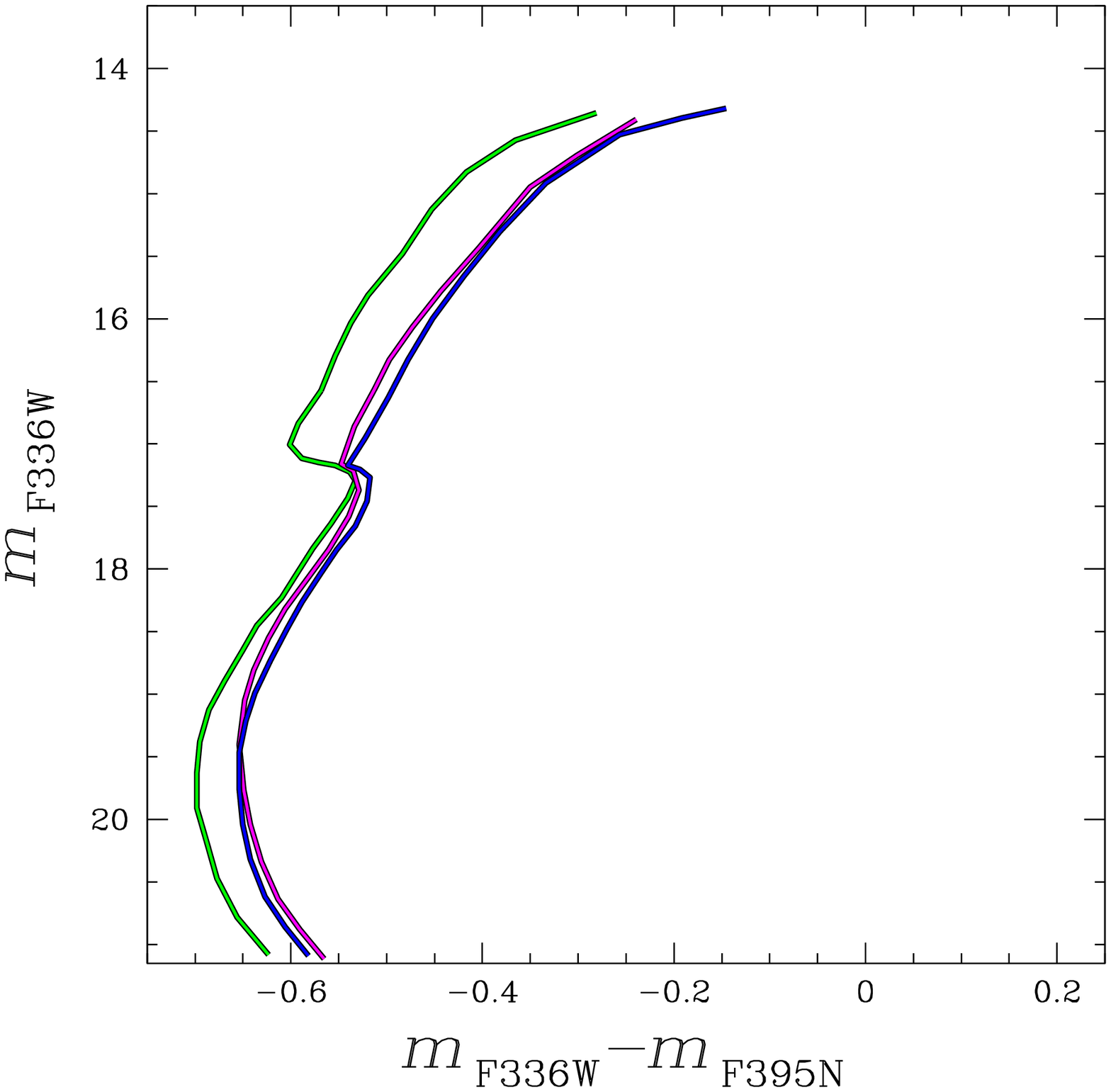}
   \caption{$m_{\rm F275W}$ versus $m_{\rm F275W}-m_{\rm F336W}$ (top) and $m_{\rm F336W}$ versus $m_{\rm F336W}-m_{\rm F395N}$ CMDs. We have colored green, magenta, and blue the three groups of stars selected in Figs.~\ref{selMS}, ~\ref{seleSGB}, and ~\ref{selRGB}. 
The fiducial lines of the three stellar populations of NGC\,6752 are shown on the right.}
\label{3pop}
\end{figure}

We analyzed \textit{HST\/} images  and ground-based catalogs obtained through
a large set of filters in order 
to identify multiple stellar populations in NGC\,6752.  We find that the MS 
of this cluster 
splits into three components, in close analogy with what we observed for its
RGB and its SGB. 
We conclude that NGC\,6752 hosts at least three stellar populations, whose
evolution can now be followed from the MS up to the RGB tip. This
result is nicely summarized in Fig.~\ref{3pop}, where we show some
representative CMDs where each population can be followed along its
evolutionary phases.\footnote{
The fiducial lines of the three stellar populations are available at this url: http://www.astro.unipd.it/globulars/
}

%__________________________________________________________________
%__________________________________________________________________

The multi-wavelength photometric data set allowed us to complement and
extend the information on the chemical composition of the different
populations available from spectroscopic chemical abundance measurements 
of a limited sample of stars. 
We calculated model atmospheres for MS and RGB stars accounting for
the available chemical composition, and demonstrated that the three
groups of stars have different helium and light-element
abundances. The most straightforward interpretation is that
Group `a', 
which contains about 25\% 
of the total number of stars, is the first stellar population,
originated in a molecular cloud with a chemical composition similar to
that of the Galactic halo, of which it shares its chemical
composition. The majority of stars represents a second (and third)
stellar populations which we named `b' (and `c') and contain about 
30\% (and 45 \%) of the cluster stars. 
They formed out of material that had been 
partly processed through first-generation stars, and are C/O poor,
N/Na/Al rich, and enhanced in helium by $\Delta Y \sim$0.01 and
$\sim$0.03, respectively.  We note that stars in each RGB sequence
exhibit a wide spread in the abundance of some light-elements
(e.g.\ N and Mg), thus suggesting that the three groups of stars
defined above are not chemically homogeneous.
Interestingly also the HB morphology of NGC\,6752 seems to be composed also by three sub-groups (Momany et al.\ 2002),
suggesting that these three populations eventually evolve in HB stars with different properties.
No radial gradient of the different stellar populations 
was detected within NGC\,6752.

Similar conclusions have been reached for
other GCs: the multiple MSs of $\omega$ Centauri and NGC\,2808 suggest
extreme helium abundances ($Y \sim 0.39$, e.g.\ Bedin et al.\ 2004,
D'Antona et al.\ 2005, Piotto et al.\ 2007, King et al.\ 2012) while
the multiple sequences of NGC\,6397 and 47 Tuc imply small helium
enhancements ($\Delta Y \sim 0.01-0.02$, Di Criscienzo et
al.\ 2010a,b, Milone et al.\ 2012a,b).  
Interestingly, while in NGC\,6397 and 47 Tuc there is evidence of two groups of stars with slight different helium content in the case of both NGC\,6752 and
NGC\,2808 we have identified at least three stellar populations.  
It is worth noting that a multimodal RGB with at least three components has been also observed in NGC\,6205 (Grundahl et al.\ 1998). A multi band study of this cluster is mandatory to estimate the helium content of its stellar populations.
Our results provide evidence that differences in helium abundance are a
quite common feature of different stellar populations in GCs.
D'Ercole et al.\ (2010, 2012) show that different combinations of
helium and CNO variations can be obtained, due to the different
modalities in which in each cluster the ejecta of stars are diluted
with primordial gas. 
Further study of color-magnitude diagrams
multiple populations may allow to constrain for each cluster the
parameters necessary to its specific model.

Finally, we discuss a potential disagreement between spectroscopic
and photometric evidence of multiple stellar populations in GCs which
needs to be solved for further progress in our understanding of GC
stellar populations.
%%%%%%%%%%%%%%%%%%%%%%%%%%%%
%__________________________________________________________________

\begin{acknowledgements}
We are grateful to the referee for many useful suggestion that significantly 
 improved the quality of the manuscript.
Support for this work has been provided by the IAC (grant 310394), 
and the Education and Science Ministry of Spain (grants AYA2007-3E3506, and AYA2010-16717).
APM acknowledges the financial support from the Australian Research 
Council through Discovery Project grant DP120100475.
AB, SC, and GP  acknowledge partial support by the ASI-INAF I/009/10/0 grant, 
and PRIN-INAF 2010.
GP acknowledge partial support by the Universita' di Padova CPDA101477 grant.
\end{acknowledgements}
\bibliographystyle{aa}

\begin{thebibliography}{}
\bibitem[Anderson(1997)]{1997PhDT.........8A} Anderson, A.~J.\ 1997,  Ph.D.~Thesis, Univ.\ of California, Berkeley

\bibitem[Anderson \& King(2006)]{2006acs..rept....1A} Anderson, J., \& King, I.~R.\ 2006, Instrument Science Report ACS 2006-01, 34 pages, 1 

\bibitem[Anderson et al.(2006)]{2006A&A...454.1029A} Anderson, J., Bedin, L.~R., Piotto, G., Yadav, R.~S., \& Bellini, A.\ 2006, \aap, 454, 1029 

\bibitem[Anderson et al.(2009)]{2009ApJ...697L..58A} Anderson, J., Piotto, G., King, I.~R., Bedin, L.~R., \& Guhathakurta, P.\ 2009, \apjl, 697, L58

\bibitem[Bedin et al.(2004)]{2004ApJ...605L.125B} Bedin, L.~R., Piotto, G., Anderson, J., Cassisi, S., King, I.~R., Momany, Y., \& Carraro, G.\ 2004, \apjl, 605, L125 

\bibitem[Bedin et al.(2005)]{2005MNRAS.357.1038B} Bedin, L.~R., Cassisi, S., Castelli, F., et al.\ 2005, \mnras, 357, 1038 

\bibitem[Bellini \& Bedin(2009)]{2009PASP..121.1419B} Bellini, A., \& Bedin, L.~R.\ 2009, \pasp, 121, 1419 

\bibitem[Bellini et al.(2010)]{2010AJ....140..631B} Bellini, A., Bedin, L.~R., Piotto, G., Milone, A.~P., Marino, A.~F., \& Villanova, S.\ 2010, \aj, 140, 631 

\bibitem[Bellini et al.(2011)]{2011PASP..123..622B} Bellini, A., Anderson, J., \& Bedin, L.~R.\ 2011, \pasp, 123, 622 

\bibitem[Bellini et al.(2013)]{2013arXiv1301.2822B} Bellini, A., Piotto, G., Milone, A.~P., et al.\ 2013, arXiv:1301.2822 

\bibitem[Briley et al.(1994)]{1994AJ....108.2183B} Briley, M.~M., Hesser, 
J.~E., Bell, R.~A., Bolte, M., \& Smith, G.~H.\ 1994, \aj, 108, 2183 

\bibitem[Briley(1997)]{1997AJ....114.1051B} Briley, M.~M.\ 1997, \aj, 114, 1051 

\bibitem[Busso et al.(2007)]{2007A&A...474..105B} Busso, G., Cassisi, S., Piotto, G., et al.\ 2007, \aap, 474, 105 

\bibitem[Cannon et al.(1998)]{1998MNRAS.298..601C} Cannon, R.~D., Croke, B.~F.~W., Bell, R.~A., Hesser, J.~E., \& Stathakis, R.~A.\ 1998, \mnras, 298, 601 

\bibitem[Carretta et al.(2005)]{2005A&A...433..597C} Carretta, E., Gratton, R.~G., Lucatello, S., Bragaglia, A., \& Bonifacio, P.\ 2005, \aap, 433, 597 

\bibitem[Carretta et al.(2007)]{2007A&A...464..927C} Carretta, E., Bragaglia, A., Gratton, R.~G., Lucatello, S., \& Momany, Y.\ 2007, \aap, 464, 927 

\bibitem[Carretta et al.(2009)]{2009A&A...505..117C} Carretta, E., et al.\ 2009, \aap, 505, 117 

\bibitem[Carretta et al.(2010)]{2010A&A...516A..55C} Carretta, E., Bragaglia, A., Gratton, R.~G., et al.\ 2010, \aap, 516, A55 

\bibitem[Carretta et al.(2011)]{2011A&A...535A.121C} Carretta, E., Bragaglia, A., Gratton, R., D'Orazi D'Orazi, V., \& Lucatello, S.\ 2011, \aap, 535, A121 

\bibitem[Carretta et al.(2012)]{2012ApJ...750L..14C} Carretta, E., Bragaglia, A., Gratton, R.~G., Lucatello, S., \& D'Orazi, V.\ 2012, \apjl, 750, L14 

\bibitem[Cassisi et al.(2009)]{2009ApJ...702.1530C} Cassisi, S., Salaris, M., Anderson, J., et al.\ 2009, \apj, 702, 1530 

\bibitem[Catelan et al.(2010)]{2010IAUS..266..281C} Catelan, M., Valcarce, A.~A.~R., \& Sweigart, A.~V.\ 2010, IAU Symposium, 266, 281 

\bibitem[Conroy(2012)]{2012ApJ...758...21C} Conroy, C.\ 2012, \apj, 758, 21 

\bibitem[Cottrell \& Da Costa(1981)]{1981ApJ...245L..79C} Cottrell, P.~L., \& Da Costa, G.~S.\ 1981, \apjl, 245, L79 

\bibitem[Dalessandro et al.(2011)]{2011MNRAS.410..694D} Dalessandro, E., Salaris, M., Ferraro, F.~R., et al.\ 2011, \mnras, 410, 694 

\bibitem[D'Antona et al.(2002)]{2002A&A...395...69D} D'Antona, F., Caloi, V., Montalb{\'a}n, J., Ventura, P., \& Gratton, R.\ 2002, \aap, 395, 69 

\bibitem[D'Antona \& Caloi(2004)]{2004ApJ...611..871D} D'Antona, F., \& Caloi, V.\ 2004, \apj, 611, 871 

\bibitem[D'Antona et al.(2005)]{2005ApJ...631..868D} D'Antona, F., Bellazzini, M., Caloi, V., Pecci, F.~F., Galleti, S., \& Rood, R.~T.\ 2005, \apj, 631, 868 

\bibitem[D'Ercole et al.(2010)]{2010MNRAS.407..854D} D'Ercole, A., D'Antona, F., Ventura, P., Vesperini, E., \& McMillan, S.~L.~W.\ 2010, \mnras, 407, 854 

\bibitem[D'Ercole et al.(2012)]{2012MNRAS.423.1521D} D'Ercole, A., D'Antona, F., Carini, R., Vesperini, E., \& Ventura, P.\ 2012, \mnras, 423, 1521 

\bibitem[Denisenkov \& Denisenkova(1989)]{1989ATsir1538...11D} Denisenkov, P.~A., \& Denisenkova, S.~N.\ 1989, Astronomicheskij Tsirkulyar, 1538, 11 

\bibitem[di Criscienzo et al.(2010)]{2010MNRAS.408..999D} di Criscienzo, M., Ventura, P., D'Antona, F., Milone, A., \& Piotto, G.\ 2010, \mnras, 408, 999 

\bibitem[di Criscienzo et al.(2010)]{2010A&A...511A..70D} di Criscienzo, M., D'Antona, F., \& Ventura, P.\ 2010, \aap, 511, A70 

\bibitem[Goudfrooij et al.(2011)]{2011ApJ...737....4G} Goudfrooij, P., Puzia, T.~H., Chandar, R., \& Kozhurina-Platais, V.\ 2011, \apj, 737, 4 

\bibitem[Gratton et al.(2001)]{2001A&A...369...87G} Gratton, R.~G., Bonifacio, P., Bragaglia, A., et al.\ 2001, \aap, 369, 87 

\bibitem[Gratton et al.(2004)]{2004ARA&A..42..385G} Gratton, R., Sneden, C., \& Carretta, E.\ 2004, \araa, 42, 385 

\bibitem[Gratton et al.(2011)]{2011A&A...534A.123G} Gratton, R.~G., Lucatello, S., Carretta, E., et al.\ 2011, \aap, 534, A123 

\bibitem[Grundahl et al.(1998)]{1998ApJ...500L.179G} Grundahl, F., Vandenberg, D.~A., \& Andersen, M.~I.\ 1998, \apjl, 500, L179 

\bibitem[Grundahl et al.(2000)]{2000LIACo..35..503G} Grundahl, F., Vandenberg, D.~A., Stetson, P.~B., Andersen, M.~I., \& Briley, M.\ 2000, Liege International Astrophysical Colloquia, 35, 503 

\bibitem[Grundahl et al.(2002)]{2002A&A...385L..14G} Grundahl, F., Briley, M., Nissen, P.~E., \& Feltzing, S.\ 2002, \aap, 385, L14 

\bibitem[Harris  (1996)]{harris96} Harris, W.\ E.  1996, AJ, 112, 1487 (December 2010 update)

\bibitem[Lind et al.(2011)]{2011A&A...527A.148L} Lind, K., Charbonnel, C., Decressin, T., et al.\ 2011, \aap, 527, A148 

\bibitem[Kraft(1979)]{1979ARA&A..17..309K} Kraft, R.~P.\ 1979, \araa, 17, 309 

\bibitem[Kraft et al.(1992)]{1992AJ....104..645K} Kraft, R.~P., Sneden, C., 
Langer, G.~E., \& Prosser, C.~F.\ 1992, \aj, 104, 645 

\bibitem[Kravtsov et al.(2011)]{2011A&A...527L...9K} Kravtsov, V., Alca{\'{\i}}no, G., Marconi, G., \& Alvarado, F.\ 2011, \aap, 527, L9 

\bibitem[Kurucz(2005)]{2005MSAIS...8...14K} Kurucz, R.~L.\ 2005, Memorie della Societa Astronomica Italiana Supplementi, 8, 14 

\bibitem[Lee et al.(2009)]{2009Natur.462..480L} Lee, J.-W., Kang, Y.-W., Lee, J., \& Lee, Y.-W.\ 2009, \nat, 462, 480 

\bibitem[Marino et al.(2008)]{2008A&A...490..625M} Marino, A.~F., Villanova, S., Piotto, G., Milone, A.~P., Momany, Y., Bedin, L.~R., \& Medling, A.~M.\ 2008, \aap, 490, 625 

\bibitem[Marino et al.(2011)]{2011ApJ...730L..16M} Marino, A.~F., Villanova, S., Milone, A.~P., Piotto, G., Lind, K., Geisler, D., \& Stetson, P.~B.\ 2011, \apjl, 730, L16 

\bibitem[Milone et al.(2008)]{2008ApJ...673..241M} Milone, A.~P., et al.\ 2008, \apj, 673, 241 

\bibitem[Milone et al.(2009)]{2009A&A...497..755M} Milone, A.~P., Bedin, L.~R., Piotto, G., \& Anderson, J.\ 2009, \aap, 497, 755 

\bibitem[Milone et al.(2010)]{2010ApJ...709.1183M} Milone, A.~P., et al.\ 2010, \apj, 709, 1183 

\bibitem[Milone et al.(2012a)]{2012ApJ...744...58M} Milone, A.~P., Piotto, G., Bedin, L.~R., et al.\ 2012, \apj, 744, 58 

\bibitem[Milone et al.(2012b)]{2012ApJ...745...27M} Milone, A.~P., Marino, A.~F., Piotto, G., et al.\ 2012, \apj, 745, 27 

\bibitem[Milone et al.(2012)]{2012ApJ...754L..34M} Milone, A.~P., Marino, A.~F., Cassisi, S., et al.\ 2012, \apjl, 754, L34 

\bibitem[Milone et al.(2012)]{2012A&A...540A..16M} Milone, A.~P., Piotto, G., Bedin, L.~R., et al.\ 2012, \aap, 540, A16 

\bibitem[Momany et al.(2002)]{2002ApJ...576L..65M} Momany, Y., Piotto, G., 
Recio-Blanco, A., et al.\ 2002, \apjl, 576, L65 

\bibitem[Momany et al.(2004)]{2004A&A...420..605M} Momany, Y., Bedin, L.~R., Cassisi, S., et al.\ 2004, \aap, 420, 605 

\bibitem[Norris \& Freeman(1979)]{1979ApJ...230L.179N} Norris, J., \& Freeman, K.~C.\ 1979, \apjl, 230, L179 

\bibitem[Norris(1981)]{1981ApJ...248..177N} Norris, J.\ 1981, \apj, 248, 
177 

\bibitem[Norris et al.(1981)]{1981ApJ...244..205N} Norris, J., Cottrell, P.~L., Freeman, K.~C., \& Da Costa, G.~S.\ 1981, \apj, 244, 205 

\bibitem[Pietrinferni et al.(2004)]{2004ApJ...612..168P} Pietrinferni, A., Cassisi, S., Salaris, M., \& Castelli, F.\ 2004, \apj, 612, 168 

\bibitem[]{} Pietrinferni, A., Cassisi, S., Salaris, M., Percival, S., \& Ferguson, J.\ W. 2009, \apj, 697, 275

\bibitem[Piotto et al.(2005)]{2005ApJ...621..777P} Piotto, G., Villanova, S., Bedin, L.~R., et al.\ 2005, \apj, 621, 777 

\bibitem[Piotto et al.(2007)]{2007ApJ...661L..53P} Piotto, G., Bedin, L.~R., Anderson, J., et al.\ 2007, \apjl, 661, L53 

\bibitem[Piotto et al.(2012)]{2012arXiv1208.1873P} Piotto, G., Milone, A.~P., Anderson, J., et al.\ 2012, arXiv:1208.1873 

\bibitem[Ram{\'{\i}}rez \& Cohen(2002)]{2002AJ....123.3277R} Ram{\'{\i}}rez, S.~V., \& Cohen, J.~G.\ 2002, \aj, 123, 3277 

\bibitem[Sbordone(2005)]{2005MSAIS...8...61S} Sbordone, L.\ 2005, Memorie della Societa Astronomica Italiana Supplementi, 8, 61 

\bibitem[Sbordone et al.(2007)]{2007IAUS..239...71S} Sbordone, L., Bonifacio, P., \& Castelli, F.\ 2007, IAU Symposium, 239, 71 

\bibitem[Sbordone et al.(2011)]{2011A&A...534A...9S} Sbordone, L., Salaris, M., Weiss, A., \& Cassisi, S.\ 2011, \aap, 534, A9 

\bibitem[Sneden et al.(1994)]{1994AJ....107.1773S} Sneden, C., Kraft, R.~P., Langer, G.~E., Prosser, C.~F., \& Shetrone, M.~D.\ 1994, \aj, 107, 1773 

\bibitem[Smith \& Norris(1993)]{1993AJ....105..173S} Smith, G.~H., \& Norris, J.~E.\ 1993, \aj, 105, 173 

\bibitem[Yong et al.(2003)]{2003A&A...402..985Y} Yong, D., Grundahl, F., Lambert, D.~L., Nissen, P.~E., \& Shetrone, M.~D.\ 2003, \aap, 402, 985 

\bibitem[Yong et al.(2005)]{2005A&A...438..875Y} Yong, D., Grundahl, F., Nissen, P.~E., Jensen, H.~R., \& Lambert, D.~L.\ 2005, \aap, 438, 875 

\bibitem[Yong et al.(2008)]{2008ApJ...684.1159Y} Yong, D., Grundahl, F., Johnson, J.~A., \& Asplund, M.\ 2008, \apj, 684, 1159 

\bibitem[Villanova et al.(2009)]{2009A&A...499..755V} Villanova, S., Piotto, G., \& Gratton, R.~G.\ 2009, \aap, 499, 755 

\end{thebibliography}

\end{document}